\documentclass[%
 reprint,
superscriptaddress,
nofootinbib,
 amsmath,amssymb,
 aps,
prx,
floatfix,
]{revtex4-2}

\usepackage{graphicx}
\usepackage{dcolumn}
\usepackage{bm}

\usepackage{lineno}
\usepackage{xcolor}
\usepackage{multirow}
\usepackage{diagbox}
\usepackage{booktabs}
\usepackage{hyperref}

\newcommand{\di}{\text{d}}
\newcommand{\pare}[1]{\left(#1\right)}
\newcommand{\parea}[1]{\left[#1\right]}
\newcommand{\avg}[1]{\left<#1\right>}
\newcommand{\abs}[1]{\left|#1\right|}
\newcommand{\gf}{G_\text{F}}

\newcommand{\ket}[1]{\left|#1\right>}
\newcommand{\bra}[1]{\left<#1\right|}

\begin{document}


\title{Superradiant Interactions \\ of the Cosmic Neutrino Background, Axions, Dark Matter, and Reactor Neutrinos}

\author{Asimina Arvanitaki}
 \email{aarvanitaki@pitp.ca}
\affiliation{Perimeter Institute for Theoretical Physics, Waterloo, ON N2L 2Y5, Canada}

\author{Savas Dimopoulos}%
 \email{savas@stanford.edu}
\affiliation{Stanford Institute for Theoretical Physics, Stanford University, Stanford, California 94305, USA}
\affiliation{Perimeter Institute for Theoretical Physics, Waterloo, ON N2L 2Y5, Canada}

\author{Marios Galanis}
 \email{mgalanis@pitp.ca}
\affiliation{Perimeter Institute for Theoretical Physics, Waterloo, ON N2L 2Y5, Canada}

\date{\today}

\begin{abstract}

In this paper we do three things. First, we outline the conditions under which the interaction rate of \emph{inelastic} processes  that change the internal state of a system of $N$ targets scales as $N^2$. This is an effect distinct from coherent \emph{elastic} scattering, but with the same scaling. Second, we compute example rates for such processes for various weakly interacting particles. Finally, we point to potential quantum observables for these processes that go beyond traditional energy exchange.

Maximal coherence in inelastic processes is achieved when the targets are placed in an equal superposition of the ground and excited states.
These coherent inelastic processes are analogous to Dicke superradiance, where cooperative effects reinforce the emission of radiation from matter, and we thus refer to them as \emph{superradiant} interactions.

We compute the superradiant interaction rates for the Cosmic Neutrino Background (C$\nu$B), dark matter scattering and absorption, and late-universe particles, such as reactor neutrinos, when the two-level system is realized by nuclear or electron spins in a magnetic field. The rates we find can be quite sizable on macroscopic yet small targets. For example, the C$\nu$B interacts with a rate of $\mathcal{O}(\text{Hz})$ when scattering off a 10~cm liquid or solid-state density spin-polarized sphere, a $\mathcal{O}(10^{21})$ enhancement compared to the incoherent inelastic contribution. For QCD axion dark matter, 
similar rates can be achieved with much smaller samples, $N \sim \mathcal{O}(10^{15})\pare{\frac{m}{2\times 10^{-8}~\text{eV}}}^{-1/2}$, where $m$ is the axion mass.

Using the Lindblad formalism for open quantum systems, we show that these superradiant interactions can manifest as a source of noise on the system. This noise is tunable however and can serve as a signature of new physics, as the energy splitting controls the momentum transfer and hence the amount of macroscopic coherence. These considerations point to new observables that go beyond traditional net energy exchange. These observables are sensitive to the \emph{sum} of the excitation and de-excitation rates -- instead of the net energy exchange rate which can be very suppressed -- and can be viewed as introducing diffusion and decoherence to the system. While we postpone to upcoming work proposing a concrete protocol that extracts these effects from a macroscopic ensemble of atoms, the effects presented in this paper may point to a new class of ultra-low threshold detectors.
\end{abstract}



\maketitle

\section{Introduction}
\label{sec:intro}

Experiments searching for dark matter (DM) candidates, the Cosmic Neutrino Background (C$\nu$B), or any other weakly interacting particle, face significant challenges due to the very weak couplings these particles have with matter. To address this, one common strategy is to use large detectors, where the interaction rates are enhanced proportionally to the number of atoms $N$. Another approach is to leverage coherent elastic interactions, where the interaction rate can be enhanced by $N^2$. These elastic interactions affect the motion of the detector's center of mass (CM) without changing its internal state.

In this paper, we investigate the coherent \textit{inelastic} interactions of weakly interacting particles with matter. A familiar example of a coherent inelastic interaction is the coherent electromagnetic (EM) radiation from an antenna composed of $N$ aligned dipoles, each with a dipole moment $p$. The antenna's total dipole moment is $N\times p$, and for wavelengths larger than the antenna's size, the dipoles radiate in phase, causing the power radiated to scale as $N^2$. This scenario is a classical analogue to the quantum mechanical phenomenon of Dicke superradiance~\cite{Dicke-superradiance}. By  time reversal, there is a corresponding enhancement in the absorption of radiation, or \emph{superabsorption}, which we aim to exploit. In direct analogy with photon superradiance, we will refer to the processes described in this paper as \emph{superradiant interactions}.

Superradiant interactions may provide a new avenue for detection by enhancing the interaction rate, as well as providing new observables associated with the excitation of the internal state of the system. 

After we outline the conditions for superradiant interactions to occur, we calculate the superradiant interaction rates for several cosmic relics: The Cosmic Neutrino Background (C$\nu$B), DM scattering and DM absorption in the form of axions, and other late-universe particles, such as reactor neutrinos,  as well as axions sourced by a different type of superradiance known as black hole superradiance. Throughout this work we calculate the rates for these particles to excite and de-excitate ensembles of two-level systems made out of nuclear or electron spins, but our analysis can be trivially generalized for any two level system. 

Furthermore, we show that these superradiant interactions can manifest as \emph{noise} in a quantum optics system. Because of this, there are observables that go beyond energy transfer, and are sensitive to the \emph{sum} of the superradiant excitation and de-excitation rates. 

Before discussing the physics of superradiant interactions, it is useful to first review the conditions under which coherence arises in elastic scattering.

\subsection{ 
Review of coherent elastic scattering}
\label{sec:form-factor-elastic-intro}

Before we outline the conditions under which the rate of inelastic processes grows quadratically with the density of the targets, it is  instructive to start with the canonical case of coherent elastic scattering~\cite{Rayleigh1}. 
Consider the C$\nu$B or a DM particle incident on a target consisting of two-state atoms (or spins). In this paper, we refer to individual targets as spins or atoms interchangeably. We focus on target particles whose mass is much larger compared to that of the particle that scatters from them. We also assume they occupy an extended volume of radius $R$ and density $n$. We assume a spherical shape for this distribution to simplify our calculation and without loss of generality.

For coherent elastic processes, the differential scattering rate from the entire structure is \cite{Rayleigh2, gans1925}:
\begin{eqnarray}
\label{eq:coh-el}
    \frac{d \Gamma_{\text{total}}(k)}{d \Omega'}=\abs{\sum_\alpha e^{i \textbf{q}\cdot \textbf{r}_\alpha}}^2\frac{d \Gamma_\alpha(k)}{d \Omega'},
\end{eqnarray}
where $\textbf{q}=\textbf{k}'-\textbf{k}$ is the momentum transfer of a particle of initial momentum $k$, and outgoing momentum $k'=k$. The sum runs over all atoms with positions $r_\alpha$ and $\frac{d \Gamma_\alpha(k)}{d \Omega'}$ is the differential scattering rate from an individual atom. When the inverse of the momentum transfer
is larger than the inter-target distance, the sum can be substituted by an integral over the volume of the scatterers yielding the well-known expression:

\begin{equation}
    \begin{split}
           \mathcal{F}(q)\equiv &\abs{\sum_\alpha e^{i \textbf{q}\cdot \textbf{r}_\alpha}}^2  \nonumber \\
   & =N + n^2 \abs{\int d^3 r_\alpha \, e^{i \textbf{q}\cdot \textbf{r}_\alpha}}^2 \\
   &= N +n^2\abs{\frac{4 \pi  (\sin (q R)-q R \cos (q R))}{q^3}}^2.
    \end{split}
    \label{eq:bisectrix}
\end{equation}

The expression above reveals that the fundamental requirement for coherent coupling to an extended target distribution is $q R<1$, for which $\mathcal{F}(q)\approx N^2$. After integrating over the outgoing particle solid angle, we recover $\Gamma_\text{total} \propto N^2$ when $k R \ll 1$, and $\Gamma_\text{total} \propto \frac{N^2}{(k R)^2}$ when $k R\gg 1$. For the scattering of light, these scalings match the well-known Rayleigh and Rayleigh-Gans regimes, respectively~\cite{Rayleigh1,Rayleigh2,gans1925,vandehulst}.

In the sections that follow, we first describe how the conditions above need to be modified to insure superradiant interaction occur (Sec.~\ref{sec:N2inel-intro}). We then calculate the superradiant excitation and de-excitation rates for the C$\nu$B (Sec.~\ref{sec:cnb}), DM (Sec.~\ref{sec:DM}), and other particles (Sec.~\ref{sec:misc}). We continue to discuss how it may be in principle possible to extract the sum or difference of these rates experimentally in Sec.~\ref{sec:dark-quantum-optics} by thinking of these superradiant processes as a source of noise in a quantum detector. We conclude with a summary and thoughts on future directions in~Sec.~\ref{sec:outlook}.


\section{Superradiant interactions}
\label{sec:N2inel-intro}

In this section we outline the conditions for particles to exhibit macroscopic coherence when they interact with large detectors \emph{and} at the same time alter the internal state of the detector. While these effects are well-known for photon absorption and emission since Dicke's work in the 50's~\cite{Dicke-superradiance}, their application to weakly coupled particles, especially in scattering, are novel. Since these new effects are tightly related to Dicke superradiance, we will be referring to these processes as \emph{superradiant interactions} for the remainder of this paper.

The formalism of the previous section needs to be modified to include the excitation or de-excitation of the target's internal degrees of freedom. The simplest system we can imagine is one consisting of two-level target particles with an energy splitting $\omega_0$, such as a spin in a magnetic field. This is also the simplest system whose energy is tunable. The scattering rate that transfers or takes away an $\omega_0$ unit of energy from the system needs to include a form factor that takes into account the initial state of the assemble $\ket{{\psi}_\text{initial}}$, the operator that acts on each constituent two-level system $ \mathcal{O}_\alpha$, and the target final state $\ket{{\psi}_\text{final}}$:
\begin{eqnarray}
\mathcal{F}(q,\omega_0)\equiv &&\sum_\text{final}\abs{\bra{{\psi}_\text{final}}\sum_\alpha e^{i \textbf{q}\cdot \textbf{r}_\alpha}\mathcal{O}_\alpha \ket{{\psi}_\text{initial}}}^2
\end{eqnarray}

It is a well-known fact that any two-level system follows an $SU(2)$ algebra and, matching the notation of \cite{Mandel_Wolf_1995}, $\mathcal{O}_\alpha={J_\pm}_\alpha$ is the raising and lowering operator for each constituent of the target system. There are three initial states that are of interest to us and we will consider in this paper (we refer the reader to App.~\ref{app:general_states} for generalized states):

\begin{itemize}
\item{All the atoms start from the ground state $\ket{g_\alpha}$. We will refer to this state as $\ket{G}=\prod_\alpha \ket{{g_\alpha}}$. This is the ground state of the whole system.}

\item{Dicke showed that a system of $N$ two level atoms that interacts with an external source through $\sum_\alpha{J_\pm}_\alpha$, has constant total angular momentum $J^2$, and it leaves on the subspace of the so called Dicke energy eigenstates spanned by 
$\ket{l,m}$, $m\in\{-l,\dots,l\}$. The ground state of a system of $N$ polarized (pseudo)spins is a Dicke state $\ket{G}=\ket{N/2,-N/2}$. The one that we will use to compute $N^2$ effects is the equatorial Dicke state, $\ket{D}\equiv\ket{N/2,0}$. There is a subtlety with considering these highly entangled states in scattering processes, because of the operator acting on them is $\sum_\alpha e^{i \textbf{q}\cdot \textbf{r}_\alpha}{J_\pm}_\alpha$, which only coincides with $\sum_\alpha{J_\pm}_\alpha$ at small $q$ transfers. This is only a small correction for us, and we address this issue in App.~\ref{app:general}.}

 \item{A state that has many similar properties to $\ket{D}$, is the equatorial coherent atomic state or product state, when each atom is in an equal superposition of excited $\ket{e_\alpha}$ and ground $\ket{g_\alpha}$ states. This state, which we denote by $\ket{P}$, can be written as:
 \begin{equation}
     \begin{split}
             \ket{P}&\equiv\prod_\alpha\frac{1}{\sqrt{2}}\pare{\ket{g_\alpha}+\ket{e_\alpha}}\\
             &=2^{-N/2}\sum_m \begin{pmatrix}
        N\\
        N/2+m
    \end{pmatrix}^{1/2}\ket{N/2,m},
    \label{eq:P_Fourier}
     \end{split}
 \end{equation}
when decomposed in the Dicke basis. The atoms evolve independently in this state, which is why this state is also known as the \emph{product state}. While this state can be routinely produced experimentally, it is neither an energy nor a spin eigenstate along any axis, since the variance $\Delta E=\omega_0^2\Delta J_z =\omega_0^2 N/4$. Nevertheless, $\ket{P}$ has the most support around the $\ket{D}$ state, and thus shares many of the same characteristics with it, as we will see below. It is worth pointing out that any state that is an $\mathcal{O}(1)$ superposition of ground and excited state will behave similarly to $\ket{P}$.
 }
\end{itemize}

\begin{figure}[ht!]
    \centering
    \includegraphics[width=0.45\textwidth]{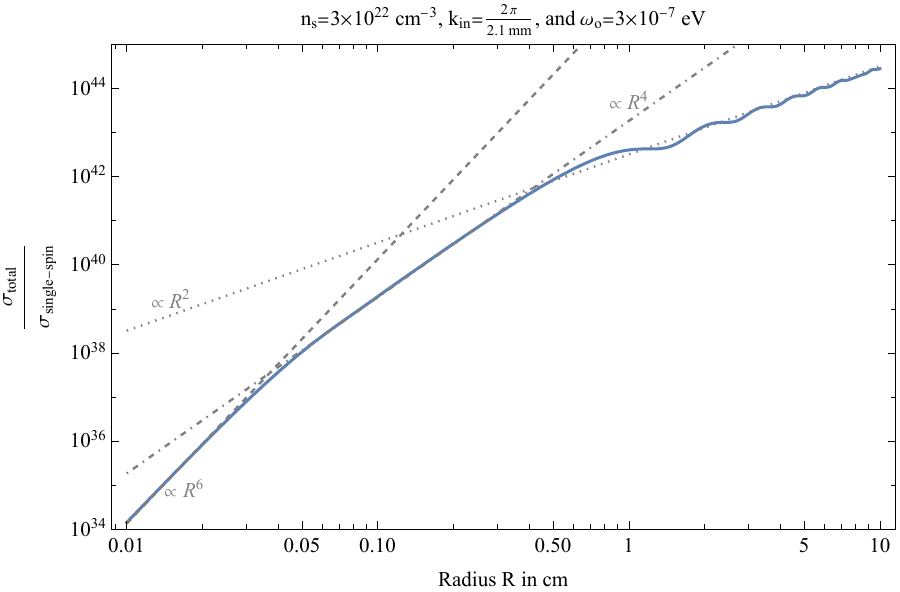}
    \caption{The ratio of the total cross-section, $\sigma_{\text{total}}$, of a neutrino wave scattering from a sphere of variable size R consisting of polarized spins with density $3\times10^{22}\text{cm}^{-3}$ to the single spin cross-section, $\sigma_{\text{single-spin}}$. The incoming neutrino momentum is set to $\frac{2 \pi}{2.1~\text{mm}}$ and the neutrino mass is taken to be $m_\nu=0.1$~eV. The excitation energy of the spins is fixed to $3\times 10^{-7}~\text{eV}$.}
    \label{fig:scaling}
\end{figure}

To find the rate of transferring energy to and from the target system through interaction with the C$\nu$B or DM, we need to sum over all the final states of the target system~\cite{Mandel_Wolf_1995}. 

We find that the form factors are given by:
\begin{eqnarray}
\label{eq:form-inel}
&&\mathcal{F}_G(q,\omega_0)=N, \\
&&\mathcal{F}_P(q,\omega_0)=\frac{1}{4}\abs{\sum_\alpha e^{i \textbf{q}\cdot \textbf{r}_\alpha}}^2+\mathcal{O}(N),
\end{eqnarray}
for an initial state $\ket{G}$ and $\ket{P}$, respectively. The form factor $\mathcal{F}_P$ is essentially the same as that of coherent elastic scattering, Eq.~\eqref{eq:bisectrix}, and should thus scale with $N^2$ for small momentum transfers. 

We immediately see that there is a non-trivial requirement for the coherent excitation of internal degrees of freedom. The targets should be placed in a superposition of ground and excited 
states in order for the coherent effects to be present. Targets starting from their ground state cannot exhibit coherent behavior. 
\begin{figure}[ht!]
    \centering
    \includegraphics[width=0.45\textwidth]{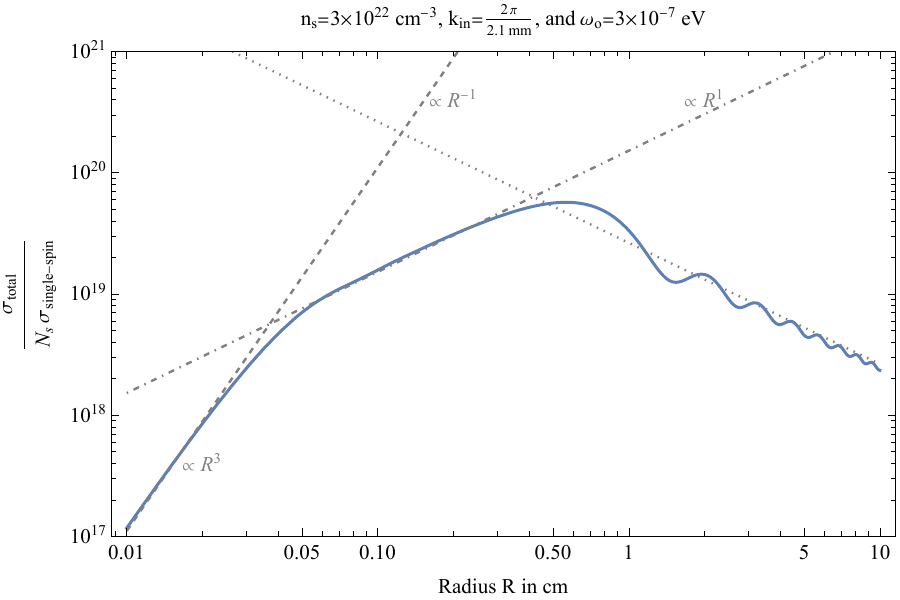}
    \caption{The ratio of the total cross-section $\sigma_\text{total}$ of a neutrino wave from a sphere of variable size to the incoherent sum of the individual spin cross-sections, $N \sigma_\text{single-spin}$, where $N$ is the total number of spins in the sphere. The spin density is $3 \times10^{22}\text{cm}^{-3}$. The incoming neutrino momentum is fixed to $\frac{2 \pi}{2.1~\text{mm}}$ and the neutrino mass is set to be $m_\nu=0.1$~eV. The excitation energy of the spins is set to $3\times 10^{-7}~\text{eV}$.}
    \label{fig:rel-scaling}
\end{figure}

There is another important difference between elastic and inelastic coherent scattering: the outgoing momentum is no longer equal to the incoming in magnitude, since energy from the scattering particle needs to be transferred to the system, for instance, in order for it to be excited. This means that for a given excitation energy $\omega_0$, there is going to be a minimum value of $q_\text{min}$. The extreme case is for example the case of absorption, which for light automatically translates to $q_\text{min}=\omega_0$. This drastically changes the scaling of the form factor when the size of the distribution $R$ becomes so large that $q_\text{min} R< 1$ can no longer be satisfied. 

The regimes of different behavior for the form factors in coherent inelastic scattering is shown in figures~\ref{fig:scaling} and \ref{fig:rel-scaling}. We have picked parameters appropriate for the C$\nu$B, so we have chosen an incoming neutrino wave with $k=\frac{2\pi}{2.1~\text{mm}}$, and mass $m_\nu=0.1$~eV. The target is an essemble of spins with density $n=3\times 10^{22}~\text{cm}^3$ and variable size. When scattering from either the $\ket{P}$ or $\ket{D}$ state, as the size $R$ of the distribution increases, the coherent interaction rate grows with $R^6$, until $R$ becomes of order $k^{-1}$. When $k R\gg 1$, full coherence is only recovered in the forward direction in a solid angle of size $(k R)^{-2}$, and the total rate scales as $R^4$. Fig.~\ref{fig:rel-scaling} shows that the overall rate still grows relative to the incoherent scattering rate.

When the momentum transfer is of order $1/R$, as imposed by coherence, the energy splitting of the spins $\omega_0$ has to satisfy:

\begin{eqnarray}
\label{eq:coherence_wmega_max}
\omega_0\leq \frac{k}{m_\nu R}=\frac{v}{R},
\end{eqnarray}
where $v$ is the velocity of the incoming particle. This implies that when $\frac{R~\omega_{0}}{v}>1$, there is not enough available momentum to excite a spin, while at the same time coupling coherently to the whole system. This is when the system transitions to the regime of $q_\text{min} R\gg 1$ of Eq.~\ref{eq:form-inel}, and the excitation rate grows as $R^2$. There is no longer a gain in the rate per atom, as Fig.~\ref{fig:rel-scaling} also shows, but the overall rate grows and it is still orders of magnitude larger than the incoherent part of the rate. 

The coherence requirement forces $\omega_0$ to be much smaller than the kinetic energy of the scattering particle. This has the natural consequence that the threshold energy of the detector is tunable, size-dependent and extremely small. This shows that coherent inelastic scattering processes may point to a new class of \emph{ultra-low threshold} particle detectors.


Moreover, another class of inelastic processes is absorption and emission or radiation. Contrary to scattering, absorption and emission of a single particle are not particle number conserving processes, and, as a result, there is always some momentum transfer set by the total energy of the particle being absorbed, i.e. $q=\omega_0$. For non-relativistic particle emission or absorption, $q$ is simply set by the momentum $k$ of the particle. In this case, coherence can only be achieved when $\omega_0 R\ll 1$, otherwise $\mathcal{F}_P$ scales like $R^2$, and there is no intermediate regime where the absorption or emission rate can grow as $R^4$.

This change in kinematics as a function of $R$ and $\omega_0$ as described above are crucial handles for signal discrimination, but this goes beyond the scope of this work. Before we go on to calculate the rates and discuss the relevant kinematics for the cases of interest,
we point out that the quantum-mechanically distinct processes of removing or adding a quantum of energy $\omega_0$ to the system happen concurrently. As we will further discuss in Sec.~\ref{sec:dark-quantum-optics}, possible observables for coherent inelastic processes scale in two possible ways with interaction rates: 

\begin{itemize}
\item{If we measure energy transfer, the relevant observable is the difference in the rate of exciting and de-exciting the system:
\begin{eqnarray}
\Gamma_\text{net}=\Gamma_+-\Gamma_-.
\end{eqnarray}}
\item{Unfortunately, in $\Gamma_\text{net}$ some or all the $N^2$ effects may cancel. In this case, it may be possible to come up with observables that are sensitive to the total interaction rate:
\begin{eqnarray}
\Gamma_\text{total}=\Gamma_++\Gamma_-.
\end{eqnarray}}
\end{itemize}

With this in mind, we are now ready to calculate sample $\Gamma_\text{net}$ and $\Gamma_\text{total}$ for several cases of interest.

\section{Superradiant scattering of the \texorpdfstring{C$\nu$B}~~from matter}
\label{sec:cnb}

The C$\nu$B is a relic of the Big Bang coming from a time when the universe was a fraction of a second old. These relic neutrinos follow a relativistic Fermi-Dirac distribution with an average temperature $T_\nu=1.95$~K (or $1.7\times 10^{-4}$~eV), and they are made from three different mass eigenstates, the heaviest has a mass experimentally constrained to be between 0.05 eV and 0.1 eV, at the most~\cite{nufit,pdg}, making it non-relativistic. These mass eigestates are linear superpositions of the three flavor states of the Standard Model. The Hamiltonian density for the interaction of a single neutrino flavor $\nu_f$ with a fermion $\psi$ receives contributions from both neutral and charged currents

\begin{equation}
    \mathcal{H}=\frac{G_\text{F}}{\sqrt{2}}\bar{\psi}\gamma^\mu\parea{g_L(1-\gamma_5)+g_R(1+\gamma_5)}\psi\,\,\bar{\nu}_{f}\gamma_\mu(1-\gamma_5)\nu_{f},
    \label{eq:hamiltonian}
\end{equation}
where $g_{L,R}$ are the left and right chiral couplings 
of $\psi$, tabulated in Table~\ref{tab:gLgR}, and $\gf=1.166\times 10^{-5}\text{GeV}^{-2}$.

For the purposes of this paper, we will focus on spin-spin interactions, and we thus pick the two-level system to consist of nuclear or electron spins in a magnetic field polarized along its direction, which we take as the $z$ direction. The energy splitting caused by the magnetic field is $\omega_0$, so that the free spin-system Hamiltonian is $H_\text{S}=\omega_0 J_z$. The interaction Hamiltonian, in the interaction picture, for the entire atomic system describing its excitation and de-excitation interactions with a flavor-eigenstate neutrino is:

\begin{equation}
\begin{split}
     H_I(t)&=\sum_\alpha H^{(\alpha)}_I(t)\\
     &=
    \frac{\gf}{2\sqrt{2}}\sum_\alpha\parea{\Sigma_-^{(\alpha)}J_+^{(\alpha)}e^{i\omega_0t}+\Sigma_+^{(\alpha)}J_-^{(\alpha)}e^{-i\omega_0 t}},
    \label{eq:neutrino-interaction-H}
\end{split}
\end{equation}
where we denote the usual Pauli operators for each spin $\alpha$ as $J_i^{(\alpha)}$, $i\in\{x,y,z\}$, while the neutrino operators $\Sigma$ are defined as $\Sigma_i^{(\alpha)}\equiv \bar{\nu}^{(\alpha)}\gamma_i(1-\gamma_5)\nu^{(\alpha)}$. In Eq.~\ref{eq:neutrino-interaction-H}, we have dropped terms that involve $J_z$ and are related to elastic scattering. We will revisit those in Sec.~\ref{sec:el_dephase}. $\nu^{(\alpha)}\equiv \nu(\mathbf{r}_\alpha,t)$ is shorthand for the neutrino quantized spinor operator evaluated at the position of the spin $\mathbf{r}_\alpha$ at time $t$.  Momentum transfer also enters through this operator, which can be expanded in plane waves and fermionic creation and annihilation operators. We have also used the shorthand notation $\mathcal{O}_\pm^{(\alpha)}\equiv \mathcal{O}_x^{(\alpha)}\pm i \mathcal{O}_y^{(\alpha)}$ for both $\mathcal{O}=J,\Sigma$.

If the momentum transfer is smaller than the inverse size of the atomic system, then we can drop the $\mathbf{r}_\alpha$-dependence of the neutrino operators. We then define the collective atomic operators $J_i\equiv \sum_\alpha J_i^{(\alpha)}$ and use the Dicke basis $\ket{N/2,m}$ to describe our system. 

In Sec.~\ref{sec:nu-elastic}, we calculate the excitation and de-excitation rates for processes where the neutrino remains in the same mass eigenstate. The cosmological evolution of the C$\nu$B allows for three different mass eigenstates to be populated, because the lifetime to decay to the lightest one is prohibitively long. This opens up the exciting possibility that neutrinos can scatter with the spins while at the same time changing their mass eigenstate. We calculate the rate for such processes in section~\ref{sec:nu-inelastic}.

\subsection{\texorpdfstring{$\nu_i\rightarrow \nu_i$}~~ scattering}
\label{sec:nu-elastic}

Here we consider the coherent inelastic scattering of a single mass eigenstate neutrino from a sample of $N$ spins, with the simultaneous excitation or de-excitation of a single spin in the sample. Summing over final helicities $h'=\pm1$ and adding the contributions from both the neutrino and the antineutrino of the same mass, the rate to deposit or remove energy into a system in the state $\ket{P}$ in the non-relativistic limit is:

\begin{equation}
\begin{split}
     \Gamma^\text{D}_{\pm}\approx\frac{\gf^2 m_\nu |u_{t,i\to i}|^2}{\pi}&\int \frac{\di^3 k}{(2\pi)^3} \frac{\sqrt{k^2\mp 2\omega_0 m_\nu}}{e^{k/T_\nu}+1}\\
     &\times\int\di\cos\theta'\mathcal{F}_P(q,\omega_0),
    \label{eq:nu_el}
\end{split}
\end{equation}
where the coefficients $|u_{t,i}|^2$ depend on the target fermion $t$ (electron, proton or neutron) and the mass eigenstates $i$ and $j$ involved in the process. $\text{D}$ denotes that this rate corresponds to Dirac neutrinos. We expect the rate for a Dicke state $\ket{D}$ to be also given by eq.~\ref{eq:nu_el}. It is given by

\begin{table}[t!]
    \centering
    \begin{tabular}{|c||c|c|c|}\hline
  \backslashbox{$g$}{$\psi$} & electron & proton & neutron \\\hline\hline
        \multirow{2}*{$g_L$} & $\nu_e$: $\frac{1}{2}+\sin^2\theta_w$ &\multirow{2}*{$\frac{1}{2}-\sin^2\theta_w$} & \multirow{2}*{$-\frac{1}{2}$} \\
       & $\nu_{\mu,\tau}$: $-\frac{1}{2}+\sin^2\theta_w$ & & \\[2pt]\hline & & & \\[-8pt]
        $g_R$ & $\sin^2\theta_w$ & $-\sin^2\theta_w$ & 0\\[3pt]\hline
    \end{tabular}
    \caption{Neutrino left and right chiral couplings with electrons, protons and neutrons in the flavor basis, as defined in Eq.~\ref{eq:hamiltonian}. Except for the left chiral coupling to electrons, all three neutrino flavors couple with the same strength to the various fermions.}
    \label{tab:gLgR}
\end{table}

\begin{figure}[t!]
    \centering
    \includegraphics[width=0.45\textwidth]{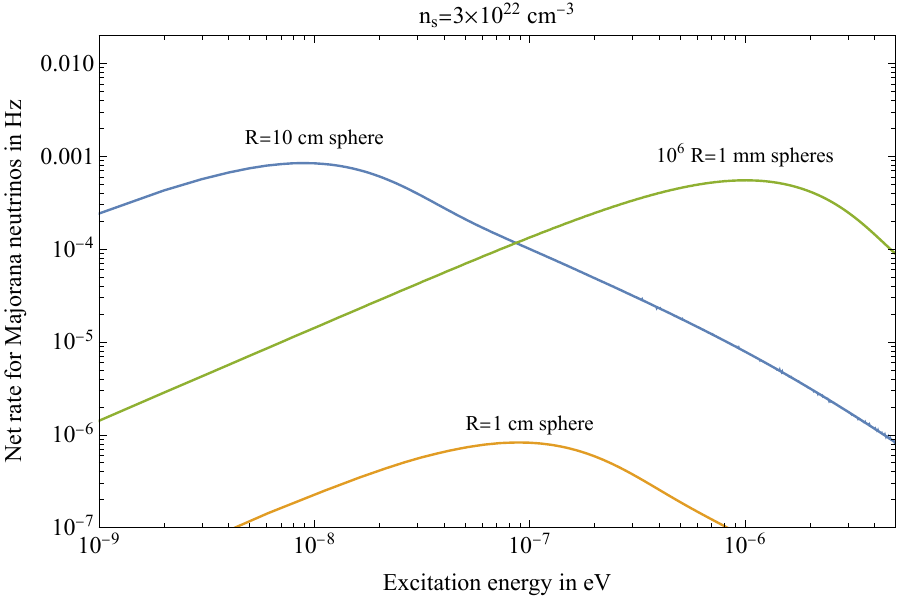}
    \caption{Net scattering rate of a Majorana C$\nu$B off of nuclear spin-polarized spheres of different sizes as a function of the spin energy splitting. For Dirac neutrinos this rate is a factor of 2 smaller. The blue curve corresponds to a sphere of radius 10~cm, the orange one to a sphere of radius 1~cm and the green one to spheres of radius 1~mm, occupying a total volume equal to that of a sphere with radius of 10~cm. The density of spins is fixed to $n_s=3\times 10^{22}~\text{cm}^{-3}$.}
    \label{fig:scatter_sphere_dif}
\end{figure}

\begin{table}[t!]
     \centering
 \begin{tabular}{|c||c|c|}\hline
    $|u|^2$ & $t=e$ & $t=p,n$\\ \hline\hline
     $1\to 1$ & 0.03 & 1/4\\
    $2\to 2$ & 0.04 & 1/4\\
     $3\to 3$ & 0.23 & 1/4 \\\hline
     $3\leftrightarrow 1$ & 0.015 & 0\\
     $3\leftrightarrow 2$ & 0.007 & 0\\
     $2\leftrightarrow 1$ & 0.2 & 0\\\hline
 \end{tabular}
 \caption{Values of the coefficients $\abs{u_{t,i\rightarrow j}}^2$ used for the rate calculations in sections~\ref{sec:nu-elastic}, and~\ref{sec:nu-inelastic}. These coefficients apply to both Dirac and Majorana fermions, as both elastic and inelastic cross-sections are not sensitive to the Majorana phases of the PMNS matrix. We took the best-fit values of the entries of $U$ from~\cite{pdg}.}
 \label{tab:u-coeff}
 \end{table}

\begin{equation}
        |u_{t,i\to j}|^2=\abs{\sum_{f=e,\mu,\tau}(g_L^{t,f}-g_R^{t,f})U_{f,j}^*U_{f,i}}^2,\quad t\in\{e,p,n\},
\end{equation}
where $U$ is the PMNS matrix, the leptonic analogue of the CKM matrix, while $g_L$, and $g_R$ are given in table~\ref{tab:gLgR}. The values of the coefficients $|u_{t,i}|^2$ are summarized in table~\ref{tab:u-coeff}. We assume three neutrinos in this work. If they are Dirac, $U$ is unitary and transitions between mass eigenstates are forbidden for proton and neutron targets because they couple only through the neutral current. If they are Majorana, $U$ is not exactly unitary. Nevertheless, the departure from unitarity is severely constrained~\cite{pdg} and so we assume $UU^\dagger=U^\dagger U=I$ in this work for both neutrino types.
The excitation and de-excitation rates of eq.~\ref{eq:nu_el} can be integrated numerically to give the total and net rate of the C$\nu$B inelastic interaction with the nuclear spin system. They are shown in figs.~\ref{fig:scatter_sphere}, and~\ref{fig:scatter_sphere_dif} for different polarized sphere sizes and configurations. If $\avg{v_\nu}$ is the average relic neutrino velocity, the total rate is maximized, when $k_\text{in}^{-1}<R< \frac{\avg{v_\nu}}{\omega_{0}}$ and it is given by:

\begin{figure}[t!]
    \centering
    \includegraphics[width=0.45\textwidth]{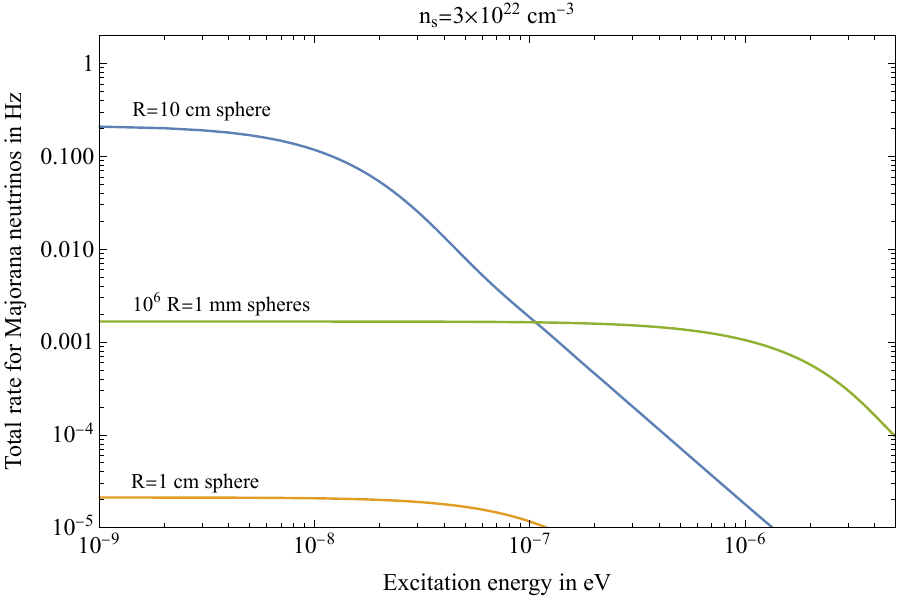}
    \caption{Total scattering rate of a Majorana C$\nu$B off of nuclear spin-polarized spheres of different sizes as a function of the spin energy splitting. For Dirac neutrinos this rate is a factor of 2 smaller. The blue curve corresponds to a sphere of radius 10~cm, the orange one to a sphere of radius 1~cm and the green one to spheres of radius 1~mm, occupying a total volume equal to that of a sphere with radius of 10~cm. The density of spins is fixed to $n_s=3\times 10^{22}~\text{cm}^{-3}$.}
    \label{fig:scatter_sphere}
\end{figure}

\begin{equation}
    \begin{split}
         \Gamma^\text{M}_\text{total}(R,m_\nu)=& \,2~\Gamma^\text{D}_\text{total}(R,m_\nu)\\
         \approx&\, 0.2~\text{Hz} \left( \frac{n_s}{3\cdot 10^{22}~\text{cm}^{-3}}\right)^2 \\&\times\left( \frac{R}{10~\text{cm}}\right)^4\left(\frac{m_\nu}{0.1~\text{eV}}\right)
    \end{split}
\end{equation}

 When $\omega_0> \frac{\avg{v_\nu}}{R}$, the rate becomes greatly reduced and eventually starts falling off exponentially when $\frac{\avg{k_\nu}^2}{2 m_\nu}< \omega_{0}$. The fact that the total interaction rate for C$\nu$B neutrinos with a 10 cm size is $\mathcal{O}$(Hz) is impressive and encouraging. It is possible when the matter is in a product or Dicke state.

In fig.~\ref{fig:scatter_sphere_dif}, we plot the net rate which is always dominated by de-excitation. It becomes maximal when $\omega_0\sim \frac{\avg{v_\nu}}{R}$, and at that optimal value of $\omega_0$ it is given by:

\begin{equation}
    \begin{split}
         \Gamma^\text{M}_\text{net}(R,m_\nu)&=2~\Gamma^\text{D}_\text{net}(R,m_\nu)\\
         &\approx 10^{-3}~\text{Hz} \left( \frac{n_s}{3\cdot 10^{22}~\text{cm}^{-3}}\right)^2 \left( \frac{R}{10~\text{cm}}\right)^3
    \end{split}
\end{equation}

The relative suppression between the total and the net rates, and the scaling of the net rate with $\omega_0$ can be understood due to difference in phase space between excitation and de-excitation, which amounts to $\frac{2 m_\nu \omega_0}{k_\nu^2}$. For the optimal $\omega_0$, this phase space difference reduces to simply $(k_\nu R)^{-1}$.

Both the net and the total rates are quite large, and, while a concrete protocol to distinguish them from the background goes beyond the scope of this paper, we revisit the detection question in section~\ref{sec:dark-quantum-optics}.

\subsection{\texorpdfstring{$\nu_\text{heavy}\leftrightarrow \nu_\text{light}$}~~scattering}
\label{sec:nu-inelastic}

The case where a neutrino mass eigenstate transmutates to another while simultaneously transferring energy to a system of atoms is very different from the case we discussed in the previous section. When $\omega_0$ is small compared to the energy released during the neutrino down-conversion, for example, the outgoing neutrino has much higher momentum than the incoming one and the minimum momentum transfer $q_\text{min}$ is generically much larger than the incoming momentum. In that case, coherence effects over a macroscopic object are suppressed. In general, the excitation and de-excitation rate for spins are given by:

\begin{equation}
\begin{split}
        \Gamma_{\pm}\approx&\frac{\gf^2 |u_{t,h(l)\to l(h)}|^2}{\pi}\\
        &\int \frac{\di^3 k}{(2\pi)^3}\int\di\cos\theta' \frac{ k'(E'-k'\cos\theta\cos\theta')}{e^{k/T_\nu}+1}\mathcal{F}_2(q,\omega_0),
    \label{eq:nu_inel}
\end{split}
\end{equation}
where $k'=\sqrt{(E_{h(l)} \mp \omega_0)^2-m_{h(l)}^2}$, and we've approximated $(E_{h(l)}+k)/E_{h(l)}\simeq 1$ inside the integral.

Unitarity of the PMNS matrix imposes that these processes are only possible when neutrinos scatter from electron spins (see table~\ref{tab:u-coeff}). The general equation above also reveals that there is a way to recover coherence, when $\omega_0$ is large enough so that $k_{\text{in}}=k_{\text{out}}$. Relic neutrinos are not monochromatic so this condition cannot be satisfied precisely, save for a very narrow range of momenta. Nevertheless, when the outgoing neutrino is also non-relativistic, which happens in the case of an inverted neutrino mass hierarchy, it is still possible to recover all or a large fraction of the coherence, when choosing:

\begin{eqnarray}
 \omega_{\text{optimal}}= m_\text{heavy}-m_\text{light}\approx\frac{\Delta m_{hl}^2}{2 m_h},
 \label{eq:nu_inel_omega_optimal}
\end{eqnarray}
where $\Delta m_{hl}^2$ is the neutrino mass-squared splitting. 
 
In this case one can easily maintain most of coherence over spheres of size $R\leq10$~cm. Even when the outgoing neutrino is quasi-relativistic, as is the case for the normal hierarchy in neutrino masses, picking $\omega_0$ appropriately close to the above value can still give a huge boost to coherence effects and maximize the interaction rate.

\begin{figure}[t!]
    \centering
    \includegraphics[width=0.45\textwidth]{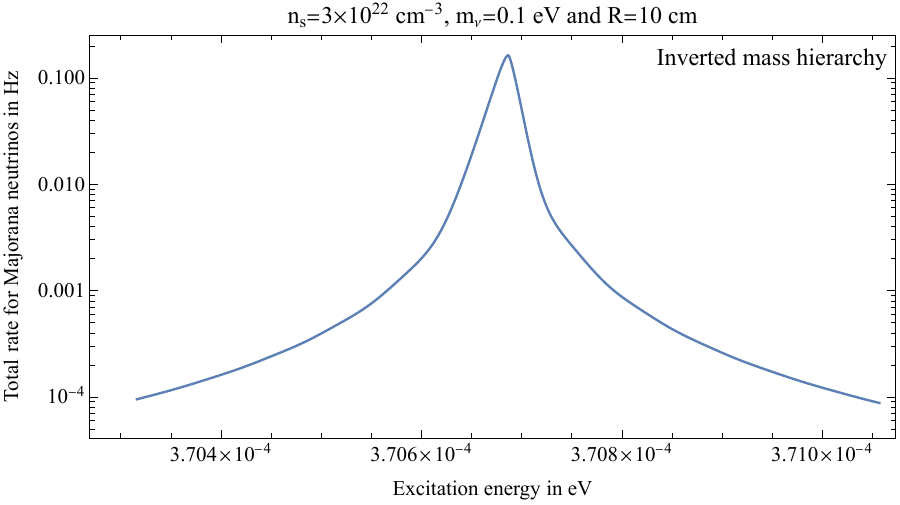}
    \caption{Total rate of neutrino downconversion or upconversion by scattering off of a spin polarized sphere of $R=10$~cm. The spin-density is fixed to $3\times 10^{22}~\text{cm}^{-3}$. The neutrino mass is fixed to 0.1 eV and an inverted mass hierarchy is assumed.}
    \label{fig:scatter_sphere_inel_inv_sum}
\end{figure}

\begin{figure}[t!]
    \centering
    \includegraphics[width=0.45\textwidth]{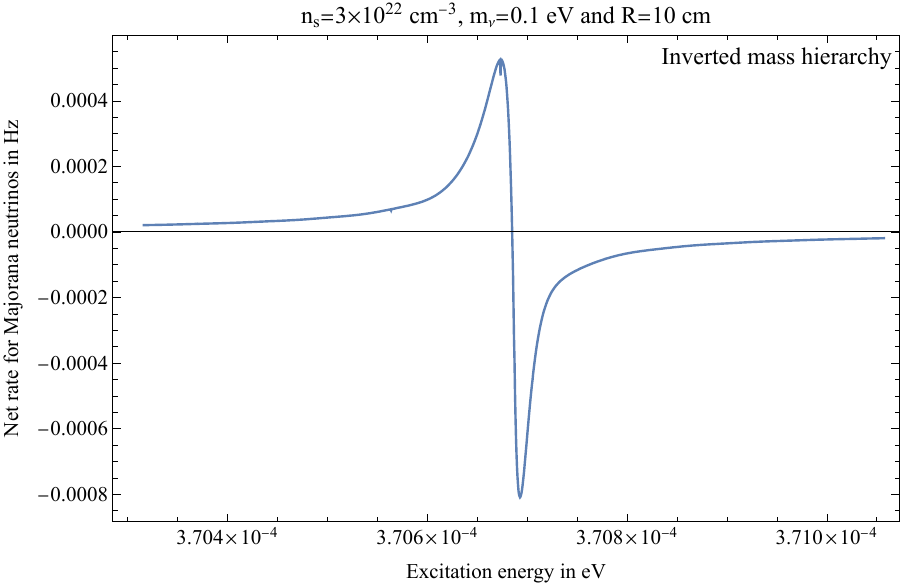}
    \caption{Net rate of neutrino downconversion by scattering off of a spin polarized sphere of $R=10$~cm. The spin-density is fixed to $3\times 10^{22}~\text{cm}^{-3}$. The neutrino mass is fixed to 0.1 eV and an inverted mass hierarchy is assumed.}
    \label{fig:scatter_sphere_inel_inv_net}
\end{figure}

This extreme dependence of $\Gamma_\pm$ on $\omega_0$ is shown in figures~\ref{fig:scatter_sphere_inel_inv_sum}, and~\ref{fig:scatter_sphere_inel_inv_net}, where we plot the total and net rates for a heavy neutrino of mass $0.1$~eV converts to a the next-lightest neutrino for inverted neutrino mass hierarchy. The total rate is maximized at $\omega_{\text{optimal}}$. At that same value, the net rate is exactly zero, because now we recover the case of elastic scattering. When $\omega_0<\omega_\text{optimal}$, the process $\nu_\text{light} \rightarrow \nu_\text{heavy}$ suffers from phase space suppression because not all modes can extract enough energy from the spin system to account from the mass deficit required in up-conversion. In this case, the process $\nu_\text{heavy}\rightarrow \nu_\text{light}$ dominates the net rate. 

Conversely, when $\omega_0> \omega_\text{optimal}$, in the process $\nu_\text{heavy}\rightarrow \nu_\text{light}$ many neutrino modes come out with momentum higher than what they came in with, coherence is lost, and the net rate is dominated by the de-excitation of the system. 

Fig.~\ref{fig:scatter_sphere_inel_inv_net} also shows the interesting feature that interaction with the C$\nu$B bath always causes a positive energy transfer from the neutrino bath to the spin system, which is unexpected. This is a direct consequence of this unusual out-of-equilibrium situation of the Universe being populated by all three neutrino mass eigenstates, instead of just the lightest one, which makes the $\nu_\text{heavy} \rightarrow \nu_\text{light}$ possible.

How close $\omega_0$ has to be chosen to $\omega_\text{optimal}$ is set by the maximum change in the average kinetic energy of the neutrino during the scattering process that still allows for coherent coupling to a structure of size $R$, i.e. $\delta \omega\sim \frac{k_\nu}{m_\nu R}$. Just as before, the maximum net rate is suppressed by roughly $(k_\nu R)^{-1}$ relative to the maximum total rate. 

\begin{figure}[t!]
    \centering
    \includegraphics[width=0.45\textwidth]{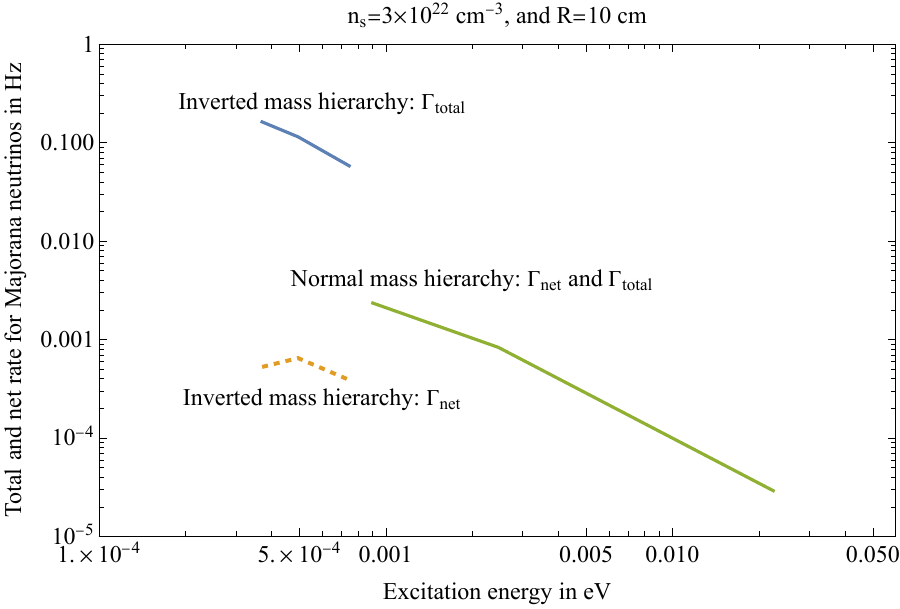}
    \caption{Total and net rate of neutrino down- or up-conversion by scattering off of a spin polarized sphere of $R=10$~cm. The spin-density is fixed to $3\times 10^{22}~\text{cm}^{-3}$. The blue curve and dashed orange curves depict the total and net rates for inverted neutrino mass hierarchy, respectively. Since the optimal resonance frequency varies a lot depending on having up- or down-conversion, the green curve depicts the total and net rate for normal neutrino mass hierarchy. For all curves, the heaviest neutrino mass varies from left to right, from 0.1~eV to 0.05~eV.}
    \label{fig:scatter_sphere_inel}
\end{figure}

Varying the heaviest neutrino mass between $0.05-0.1$~eV and considering both the case of inverted and normal mass hierarchies for neutrinos, we plot in figure~\ref{fig:scatter_sphere_inel} the total and net rates of the C$\nu$B scattering from a 10 cm electron spin polarized sphere as a function of $\omega_\text{optimal}$. For the inverted mass hierarchy case, the mass difference between  $m_1$ and $m_2$ is small enough that picking $\omega_\text{optimal}$ from eq.~\ref{eq:nu_inel_omega_optimal} insures coherence for many momentum modes.

In the case of normal hierarchy, the mass splitting between $m_3$ and $m_2$ is so large that $\omega_\text{optimal}$ can deviate significantly from eq.~\ref{eq:nu_inel_omega_optimal}. This makes the overall total rates significantly smaller than the previous case, and it makes the value of $\omega_\text{optimal}$ significantly different between the excitation and de-excitation processes that in essence $\abs{\Gamma_\text{total}}=\abs{\Gamma_\text{net}}$. The maximum excitation and de-excitation rates occur at sufficiently different values of $\omega_0$ that there is no real cancellation between the two. Even though the rates are routinely smaller than the case studied in Sec.~\ref{sec:nu-elastic}, they can still in principle be larger than 1 interaction per day. 

It is worth pointing out that the natural linewidth of the two-level system has to be small enough so that the experiment can distinguish between the up-conversion and down-conversion processes.

Finally, we point out that when $\omega_{\text{optimal}}$ is larger than $ 10^{-4}-10^{-3}~\text{eV}$, we can no longer rely on electron spins in a magnetic field for our two-level system. Above these energies, atomic transitions between levels split by spin-orbit coupling effects are available. This most likely also suppresses the maximum number densities possible, which will impact the rate. Since the purpose of this section is not an experimental proposal but a way to outline the dynamics of coherence and the relevant effects that maximize the interaction rates, we will not work out this situation in detail. However, it is worth noting that this unusual possibility of neutrino transmutation for the C$\nu$B offers a unique potential signature of the C$\nu$B signal.


\section{Dark Matter superradiant interactions}
\label{sec:DM}

The C$\nu$B is a relic of the Big Bang whose properties are very well predicted in the Standard Model framework, taking into account $\Lambda_\text{CDM}$ cosmology, making it an obvious experimental target. Another cosmological relic which is known to exist but its properties are not so well constrained is of course Dark Matter (DM). In this section, we will consider two possibilities for DM. Firstly, DM that can scatter from a two-level system, in direct analogy with the C$\nu$B. We then move on to discuss DM candidates that can be absorbed by the two-level system:~axions and dark photons.


\subsection{Dark Matter scattering}
\label{sec:DMsc}

 For concreteness and in direct analogy with the C$\nu$B , we focus on fermionic or bosonic DM that couples to spins through the operators:

\begin{align}
\label{eq:operators}
    \mathcal{L}&\supset \frac{1}{\Lambda^2}\bar{\psi}\gamma_\mu\gamma^5 \psi\,\bar{N}\gamma^\mu\gamma^5 N\\
    \mathcal{L}&\supset \frac{1}{\Lambda^2}\frac{i}{2}\pare{\phi^\dagger\partial_\mu\phi -\phi\partial_\mu\phi ^\dagger}\,\bar{N}\gamma^\mu\gamma^5 N
\end{align}

The above operators can be used to calculate the scattering rates of DM from an extended distribution of targets. The rate for fermion DM to scatter from $\ket{D}$ to the $\ket{N/2,\pm 1}$ states, or exchange one quantum of energy $\omega_0$ with the $\ket{P}$ state is:

\begin{equation}
\begin{split}
    \Gamma_\pm=&\frac{2\rho_\text{DM}}{m_\text{DM}^3\Lambda^4}\int \di^3 k\pare{\frac{1}{\pi v_0^2}}^{3/2}e^{-\frac{k^2}{m^2v_0^2}}\\
    &\times\int d\cos\theta'\frac{\sqrt{k^2\mp 2 m \omega_0}}{2 \pi}\mathcal{F}_P(q,\omega_0),
    \label{eq:BlueDragon-DM}
\end{split}
\end{equation}

where $\rho_\text{DM}\approx 0.3~\text{GeV/cm}^3$ is the local DM density, and $v_0\approx235~\text{km/s}$ is the velocity dispersion of virialized DM in the galaxy. The form factor $\mathcal{F}_P(q,\omega_0)$ has been defined in eq.~\ref{eq:form-inel}.

For bosonic Dark Matter candidates, the rates can be similarly calculated. We find that there is an additional velocity suppression due to angular momentum conservation. The corresponding excitation and de-excitation rates scattering from the $\ket{P}$ state is given by:

\begin{equation}
    \begin{split}
        \Gamma_{\pm}=&\frac{1}{96 \pi\Lambda^4}\int\frac{\di^3 k}{(2\pi)^3}\\
        &\times\int\di \cos \theta'k' \pare{1+\avg{n_{k'}}}\avg{n_k}\frac{k^2+k'^2}{\sqrt{k'^2+m^2}}\mathcal{F}_P(q),
    \end{split}
\end{equation}
where we have assumed that both $\phi$ and $\phi^\dagger$ are equally populated so that the mean occupation number of each is $\avg{n_k}=(2\pi)^3(\rho_\text{DM}/2) m^{-4}(\pi v_0^2)^{-3/2}e^{-k^2/(m^2v_0^2)}$, and $k'=\sqrt{k^2\mp 2 m \omega_0}$ by energy conservation, in the non-relativistic limit. 

\begin{figure}[t!]
    \centering
    \includegraphics[width=0.45\textwidth]{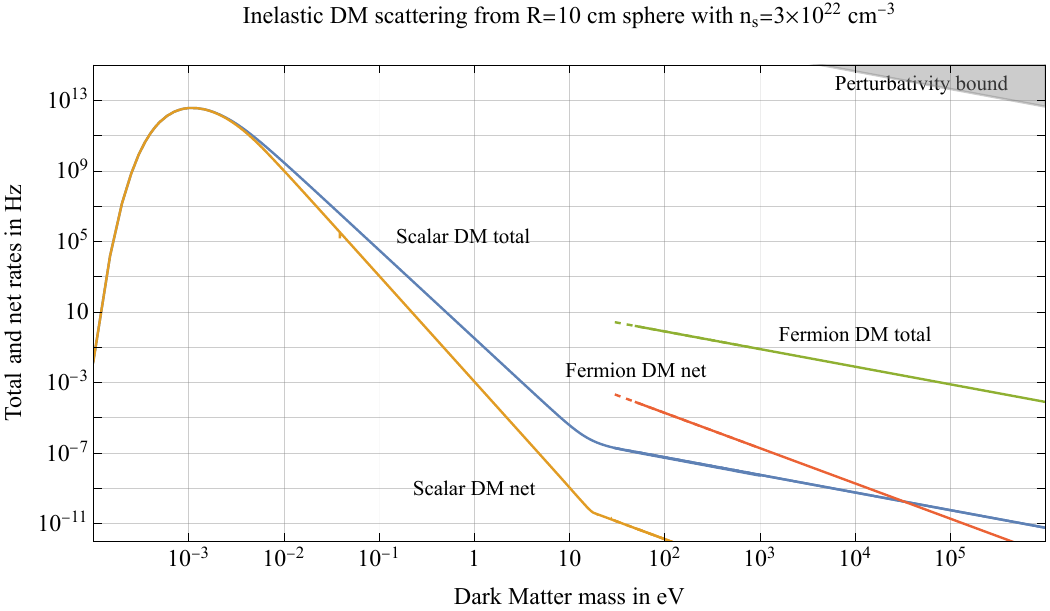}
    \caption{Upper bound on the total and net rate of fermion and scalar DM scattering from a spin-polarized R=10 cm sphere with $n_s=3\times 10^{22}~\text{cm}^{-3}$, as a function of the DM mass. The gray shaded shows the region where the rate corresponds to a larger than geometric cross-section and perturbation theory breaks down. The excitation energy is fixed to $\omega_0=1.5\times 10^{-9}~\text{eV}$.}
    \label{fig:DMscatter}
\end{figure}

In fig.~\ref{fig:DMscatter}, we show the maximum astrophysically allowed rates for scalar or fermion DM interacting inelastically with a $R=10$~cm size nuclear spin-polarized sphere. Hadrophilic DM is constrained from its production in the supernova 1987a event~\cite{supernova-DM-bounds}, and what we plot should be considered an upper bound to the total rate for the example considered here. For large masses, the total rate drops like the DM mass as expected, while the net rate gets an additional suppression of $\frac{\lambda_{DM}}{R}$. Scalar DM scattering rates get suppressed by $v_{DM}^2$, but at the same time get a boost from bose-enhancement factors, when the DM mass becomes lighter than $\sim 10$~eV. This is an example of the rate being enhanced by the square of the DM particle density and well as $N^2$! As the scalar DM mass becomes smaller, the rate eventually gets suppressed when $\frac{\langle k^2 \rangle}{2 m}\leq \omega_0$, as the DM does not have enough energy to excite the spins, or conversely, the DM momentum is too high in de-excitation processes for it to couple coherently to the entire target and there is no longer a boost from the DM large occupation number. It is also worth point out that the maximum $\omega_0$ for which full coherent excitation of an object of size $R$ can be achieved is different from that for the C$\nu$B because the DM velocity distribution in the galaxy is vastly different than that of the mostly unclustered relic neutrinos.

In principle, one can imagine scenarios where DM comprises of two components, where scattering allows transitions from one component to the other, similar to the behavior of the C$\nu$B. This can happen for example, if DM has structure similar to atoms. Devising such a contrived model, where DM does what the C$\nu$B is naturally set up to do, goes beyond the scope of this paper.

Finally, the DM rates discussed above can be much enhanced in the presence of a light mediator in the DM interactions with ordinary matter. In the case of a scalar mediator, in matter effects on the scalar's mass and potential can be very important. A light dark photon mediator may be a more viable option, but in that case, DM is essentially millicharged and its behavior in the galaxy is not well understood~\cite{millichargedDM}. It is not clear that millicharged DM is a viable DM candidate. We thus conclude that the case of a light mediator enhancing the scattering rate is better left to future work.


\subsection{Dark Matter absorption and emission: Axions and Dark Photons}
\label{sec:DMabs}

Coherence in inelastic processes is not restricted to scattering and similarly occurs for emission and absorption of bosonic particles. The obvious example is Dicke superradiance, which is coherent \emph{spontaneous} emission of radiation; both from $\ket{D}$ and from the noisy $\ket{P}$, the probability to emit a photon with a wavelength longer than the size of the system scales as $N^2$~\cite{Dicke-superradiance}. As we will see here, the situation is more subtle than scattering, but similar scalings can persist for the absorption and \emph{stimulated} emission of bosonic dark matter and, in particular, of axions and dark photons.

\subsubsection{Axion Dark Matter}
\label{sec:AxionDMabs}

The QCD axion and axion-like particles can have couplings to SM fermions $\psi$ linear in the axion field $\mathcal{L}\supset g \partial_\mu a\bar{\psi}\gamma^\mu\gamma^5\psi$, where the coupling is related to the axion decay constant $g\propto f_a^{-1}$, with a numerical coefficient that is model-dependent~\cite{krauss1985spin,Wilczek-spin-forces,pdg}. Intuition from the interaction of light with matter can carry almost verbatim to the interaction of axions with matter.

\begin{figure}[t!]
    \centering
    \includegraphics[width=0.45\textwidth]{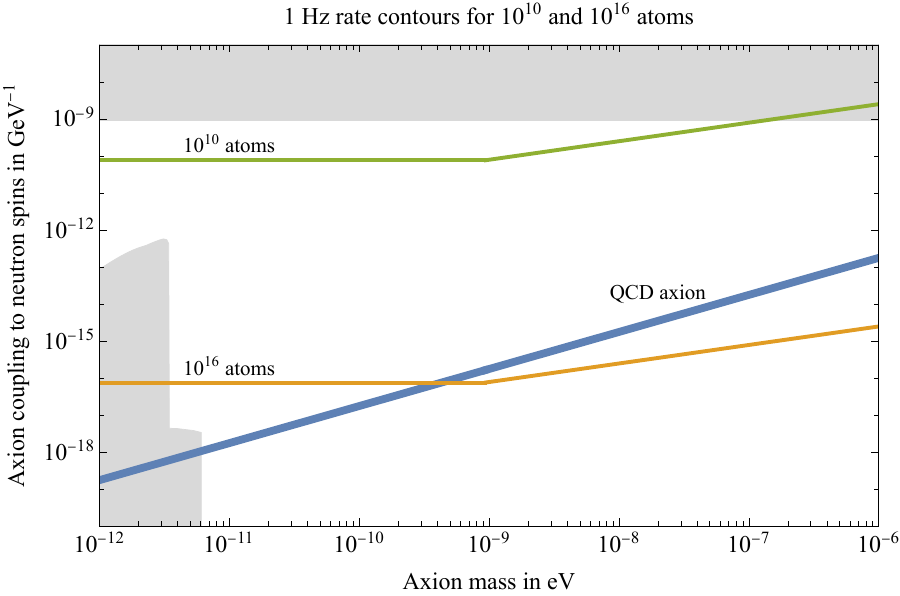}
    \caption{Axion DM 1 Hz total rate contours in the axion mass vs coupling plane for different values of the number of atoms. The gray shaded is excluded by atrophysical observations~\cite{AxionLimits, BHSR-self-interactions}. The break on the contours coincides with the axion coherence time equal to the interogation time of the experiment, which we set to 1~sec. The blue line shows the expected QCD axion coupling vs mass scaling. We also assume that the axion -- neutron spin coupling is exactly $f_a^{-1}$.}
    \label{fig:axionDM}
\end{figure}

The Hamiltonian for the excitation and de-excitation interactions of an axion with an electron or nuclear spin, whose splitting is $\omega_0$, is:

\begin{eqnarray}
    H\approx B_- J_+ e^{i\omega_0 t}+B_+ J_- e^{-i\omega_0 t},
\end{eqnarray}
where we have assumed that the axion Compton wavelength is much larger than the atomic system, so that we can again use the collective spin operators. The ``magnetic field'' $B_\pm\equiv g (\partial_x  \pm i\partial_y) a$, and the stochastic axion field is given by:

\begin{equation}
     a(\mathbf{x},t)=\frac{1}{L^{3/2}}\sum_\mathbf{k} \frac{1}{\sqrt{2m}}\tilde{a}(\mathbf{k}) e^{-i\pare{m+\frac{k^2}{2m}}t+i
    \mathbf{k}\cdot\mathbf{x}+i\phi_k}+\text{c.c.},
    \label{eq:axion}
\end{equation}
where $L^3$ is the quantization volume, the spectrum $|\tilde{a}(\mathbf{k})|^2\propto\exp\parea{-k^2/(m^2v_0^2)}$ is normalized appropriately to the local DM density, and the phases $\phi_k$ are random and uniformly distributed. It is worth pointing out that this classical description can be substituted by a second-quantized treatment similar to the one for thermal light~\cite{Mandel_Wolf_1995}. However, just like thermal light, the two are physically equivalent in what we are discussing in this work, up to spontaneous emission of axions. 

What matters for the proper treatment of axion DM is the two-point temporal correlation function at the same spatial position. For the axion gradient coupling above, the relevant quantity is:

\begin{equation}
\begin{split}
\label{eq:axion-corr}
    g^{(1)}_\pm(\tau)&\equiv \frac{1}{f_a^2}\avg{\parea{\nabla a (t+\tau)}_\pm\parea{\nabla a(t)}_\mp}\\
    &=\frac{\rho_\text{DM}v_0^2\cos\pare{m\tau+\frac{5}{2}\arctan\frac{mv_0^2\tau}{2}}}{f_a^2\parea{1+\pare{\frac{mv_0^2\tau}{2}}^2}^{5/4}},
\end{split}
\end{equation}
where we defined the coherence time as $\tau_c\equiv (mv_0^2)^{-1}$.

There are several similarities to thermal light. In the large $\tau$ limit, eq.~\eqref{eq:axion-corr} shows that correlations decay, so that the system loses memory of each interaction event and thus exhibits \emph{Markovian} behavior. When $\tau\lesssim \tau_c$, the stochastic axion exhibits correlations, so that the system's response exhibits \emph{non-Markovian} behavior. This regime also occurs for light, when the temperature is low enough that the photon bath fluctuations have coherence time comparable to other experimental timescales.

The Markovian axion DM regime is quite similar to scattering, where coherence times are extremely short. Similarly to sec.~\ref{sec:cnb}, we wish to compute the axion absorption and emission rates on the $\ket{P}$ or $\ket{D}$ states. 
Let us first carry out the computations for a system prepared in $\ket{D}$. These are optimized by choosing $\omega_0=m +\frac{k_0^2}{2m}$ with $k_0\equiv\sqrt{\frac{3}{2}} m v_0$ and are:

\begin{eqnarray}
\begin{aligned}
   \Gamma_+=\Gamma_- &= \frac{N}{2}\pare{\frac{N}{2}+1}\frac{k_0^3}{3\pi f_a^2} |\tilde{a}(k_0)|^2 \\
    &\approx 1.9\frac{N}{2}\pare{\frac{N}{2}+1}\frac{\rho_\text{DM}}{f_a^2 m}\\
    &\approx 1\text{Hz}\pare{\frac{N}{10^{15}}}^2\pare{\frac{m}{2\cdot 10^{-8}\,\text{eV}}},\quad \\
    &\quad\text{for the QCD axion}.
\end{aligned}
\label{eq:rates_axion_Dicke}
\end{eqnarray}

We emphasize here the appearance of the occupation number per momentum mode, $|\tilde{a}(k_0)|^2$, evaluated at the resonant momentum. This is a manifestation of energy conservation, which picks out only the resonant axions. The rate $\gamma_1\equiv k_0^3/(3\pi f_a^2)$ is that of spontaneous axion emission from a single spin, making Bose-enhancement manifest.

Since the rates of excitation and de-excitation are exactly equal, the net rate is obviously zero, even for a Dicke state, up to spontaneous emission  effects, which we ignore in the equation for $\Gamma_-$. We emphasize that this behavior is markedly different from the case of scattering, where there is always a non-zero $\avg{J_z}$. Despite this, the axion still flips individuals spins with a rate given by Eq.~\eqref{eq:rates_axion_Dicke}, but at a random direction each time. The appropriate observable then is not a drift of the \emph{mean} spin of the system, but rather a \emph{diffusion}, namely an increase in the variance of the spin. Such observables are most easily derived in the framework of open quantum systems, which we study in Sec.~\ref{sec:dark-quantum-optics}.

In the non-Markovian regime the axion's phase and direction are a priori unknown, so there is an equal probability that the spin system will precess along the $+z$ and $-z$ directions. For a system prepared in $\ket{D}$, we can compute the probabilities $p_\pm$ to find it in the states $\ket{N/2,\pm1}$, respectively,  after time $t$:

\begin{equation}
    \begin{split}
        p_+=p_-&\approx\frac{N}{2}\pare{\frac{N}{2}+1}\frac{2\rho_\text{DM} v_0^2}{f_a^2}\frac{t^2}{2}\\
        &\approx 40\% \pare{\frac{N}{ 10^{16}}}^2\pare{\frac{8\cdot 10^{15}\text{GeV}}{f_a}}^2\pare{\frac{t}{1\text{s}}}^2,
        \label{eq:Rabi_axion_Dicke}
    \end{split}
\end{equation}
where we identify the Rabi frequency $\Omega^2\equiv 2\rho_\text{DM} v_0^2/f_a^2$. We have chosen GUT-scale $f_a$ so that it corresponds to a QCD axion whose coherence time is $1$s. The $t^2$ scaling of these probabilities, contrary to the $\propto t$ scaling for scattering or emission/absorption, is a clear manifestation of the axion's coherence. We explain some of the subtleties of the non-Markovian regime and its connection to Rabi oscillation in App.~\ref{app:product_classical_axion}.

Eqs.~\ref{eq:rates_axion_Dicke}, and~\ref{eq:Rabi_axion_Dicke} are both summarized in fig.~\ref{fig:axionDM}, which shows the effective 1 Hz contour total rates for $10^{10}$ and $10^{16}$ atoms. Below axion masses of $10^{-9}$~eV, the interrogation time of the experiment, which we take to be 1~sec, becomes shorter than the coherence time of the axion, and the effective rate plotted is the $\mathcal{O}(1)$ probability contour for eq.~\ref{eq:Rabi_axion_Dicke}. We stress that these contours should not be confused as possible reach of an experimental setup, as this goes beyond the scope of this paper.

\subsubsection{Dark Photon Dark Matter}
\label{sec:darkphotonDM}

Dark photons are cousins of ordinary photons, and as such, they can cause spin flips through the magnetic moment coupling. Dark photon DM that kinetically mixes with photons through $\epsilon F_\text{dark} F_\text{EM}$ manifests as a magnetic field of size~\cite{Darkphoton}:

\begin{displaymath}
B_\text{dark} \sim\epsilon \sqrt{\rho_\text{DM}}\left\{ \begin{array}{ll}
1 & \textrm{if $m R> 1$}\\
(m R) & \textrm{if $m R< 1$}
\end{array} \right.
\end{displaymath}
The dark photon effective magnetic field is affected by the presence of a shield, and $R$ denotes its size, which we fix to 1~m. Its direction end exact magnitude is determined by the shape of the shield, and the above equation should be considered as an order of magnitude estimate.

The interaction strength with a system of spins, now depends on the magnitude of the gyromagnetic ratio, $g_f$, which is about  $\mathcal{O}(10^3)$ larger for electron spins than nuclear spins.

We can now calculate the corresponding excitation and de-excitation rates of the $\ket{D}$ state in the markovian regime:

\begin{equation}
    \begin{split}
        \Gamma_+=\Gamma_-\sim&\, \frac{N}{2}\pare{\frac{N}{2}+1}\epsilon^2 g_f^2\frac{\rho_\text{DM}\times \textrm{Min}\{ (m R)^2,1\}}{m v_0^2}\\
    \approx&\, 1\text{Hz}\pare{\frac{N}{10^{15}}}^2\pare{\frac{\epsilon}{5 \cdot 10^{-17}}}^2\\
    &\times\pare{\frac{g_f}{g_N}}^2\pare{\frac{m}{2\cdot 10^{-8}\text{eV}}},
\label{eq:rates_darkphoton_Dicke}
    \end{split}
\end{equation}
where $g_N$ is the gyromagnetic ratio of the neutron.

\begin{figure}[t!]
    \centering
    \includegraphics[width=0.45\textwidth]{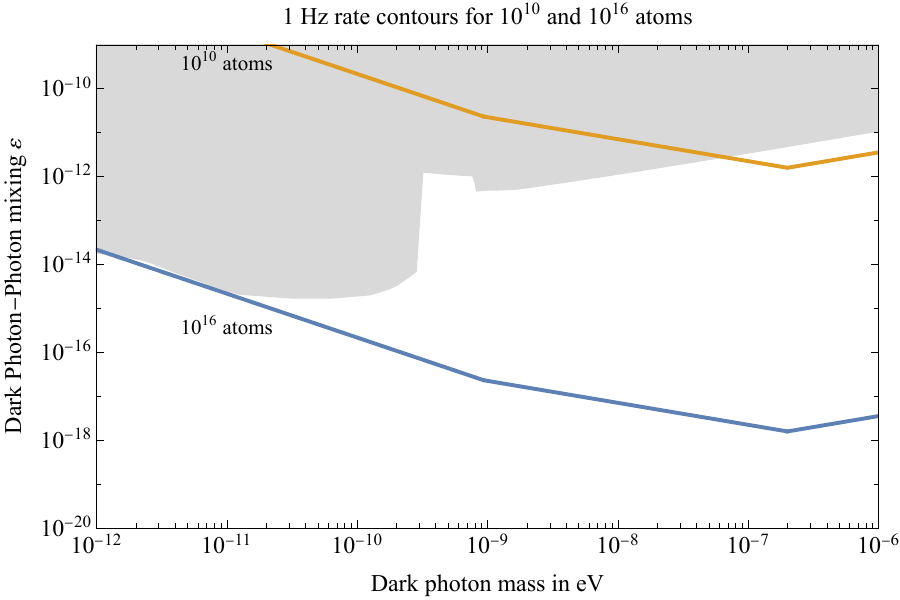}
    \caption{Dark photon DM 1 Hz total rate contours in the dark photon mass vs photon kinetic mixing plane for different values of the number of atoms. The gray shaded is excluded by a variety of astrophysical data~\cite{Caputo:2021eaa}. The break on the contours at roughly $10^{-9}$~eV mass coincides with the dark photon coherence time equal to the interogation time of the experiment, which we set to 1~sec. The break at around $10^{-7}$~eV is present because of the assumption of a shield of size $1$~m. See text for more details.}
    \label{fig:darkphotonDM}
\end{figure}

In the non-markovian regime following the prescription outlined in~\ref{sec:DMabs} we find:

\begin{equation}
    \begin{split}
        p_+=p_-\sim&\,\frac{N}{2}\pare{\frac{N}{2}+1}\epsilon^2 g_f^2 (m R)^2 t^2\\
        \approx&\, 40\% \pare{\frac{N}{ 10^{16}}}\pare{\frac{\epsilon}{5 \cdot 10^{-17}}}^2\\
        &\times\pare{\frac{g_f}{g_N}}^2\pare{\frac{m}{7\cdot 10^{-10}~\text{eV}}}^2\pare{\frac{t}{1\text{s}}}^2.
        \label{eq:Rabi_darkphoton_Dicke}
    \end{split}
\end{equation}

The estimates of eqs.~\ref{eq:rates_darkphoton_Dicke}, and~\ref{eq:Rabi_darkphoton_Dicke} are summarized in fig.~\ref{fig:darkphotonDM}, where we present two different contours of total absorption and emission rate for dark photon DM that kinetically mixes with the ordinary photon. Just like the axion case, we fix the interrogation time of the experiment to 1~sec, so the transition between the regime of validity of~eqs.~\ref{eq:rates_darkphoton_Dicke}, and~\ref{eq:Rabi_darkphoton_Dicke} happens at $10^{-9}$~eV. The presence of the shield no longer affects the dark photon DM total rate above masses of roughly $10^{-7}$~eV.

To conclude, it is obvious that the possibility of superradiant absorption or emission of axions or dark photons results in exceedingly large rates on small sample sizes. Such huge rates though are experimentally relevant if they can be distinguished from the background. Furthermore, traditional observables that rely on energy transfer are not appropriate because these excitation and de-excitation rates for axions exactly cancel up to the spontaneous emission rate that is independent from the local DM density. As we plan to discuss in section~\ref{sec:dark-quantum-optics}, there may be observables that are sensitive to the sum of these rates, or can design protocols such as the one proposed in~\cite{Higgins_2014} that manage to isolate superradiant absoprtion from superradiant emission. To conclude, we believe that the numbers calculated in this section, as outlandish as they may seem at first, may be experimentally relevant.



\section{Late-universe particles}
\label{sec:misc}

The superradiant effects discussed may prove advantageous not only for the detection of cosmological relics, but for other weakly interacting particles as well. In this section, we will discuss two cases: MeV neutrino sources, reactor, solar, and nuclear weapon neutrinos, as well as axions sourced by black hole superradiance.

\subsection{Solar, reactor, and nuclear weapon neutrinos}
\label{sec:nuclear-nus}

 The C$\nu$B is unique not only because it is a Big Bang relic, but also because it is the only sizeable non-relativistic source of neutrinos that we have in the Cosmos. Neutrinos that are produced in stars, in reactors, and in cosmic rays, are the product of high energy processes, and as a consequence, ultra-relativistic. In this section, we estimate the superradiant interaction rates of such high-energy neutrino sources, with focus on those with the highest fluxes after the C$\nu$B: solar, reactor, and nuclear weapon neutrinos. 

 These neutrinos have characteristic energies between $1-10$~MeV.
 One may naively think that it is impossible to gain long-range coherence over the entire solid, since the wavelength of these neutrinos is much smaller than the inter-atomic spacing in solids. As we saw previously, the quantity that determines this long-range coherence is the momentum transfer $q$ and not the wavelength. So, even for MeV momenta the coherent interaction cross-section can be sizeable~\cite{weiss-neutron-SANS}.

The intrinsic polarization and directionality of these neutrino sources gives a signal with a non-trivial angular dependence. A right-helical antineutrino excites the atomic system with a matrix element proportional to $\cos\frac{\theta}{2}\sin\frac{\theta'}{2}$, while it de-excites with $\sin\frac{\theta}{2}\cos\frac{\theta'}{2}$, where $\theta$ and $\theta'$ are the angles of the incoming and outgoing momenta with respect to the spin-polarization axis of the spin system, assuming that helicity does not change (the amplitude to change the helicity of a relativistic neutrino is suppressed by $m_\nu/E_\nu$). Similarly, a left-helical neutrino excites and de-excites the atomic system with matrix elements proportional to $\sin\frac{\theta}{2}\cos\frac{\theta'}{2}$ and $\cos\frac{\theta}{2}\sin\frac{\theta'}{2}$, respectively. One sees that excitation and de-excitation can be $\mathcal{O}(1)$ different if the atomic system is orientated parallel or anti-parallel to the neutrino beam direction, giving even \emph{net excitation} of the spin-system. At the same time, however, the matrix element is phase-space suppressed in the forward direction where coherent coupling is maximal. Furthermore, since the beam cannot be collimated to within $|\theta-\theta'|\sim\lambda/R$ for MeV neutrinos, excitation and de-excitation will not be much different in realistic conditions. On the other hand, if the spins are oriented perpendicularly to the neutrino beam direction, i.e. $\theta\approx \pi/4$, the interaction rate is maximal in the forward direction $\theta'\approx \theta$, but the excitation and de-excitation rates are equal, again suppressing observables related to net energy exchange. In the end, orientation effects can change the rates to $\mathcal{O}(1)$, potentially aiding in signal discrimination, but cannot help parametrically.
 
 The average neutrino flux from the sun is $10^{11}~\text{cm}^{-2}~\text{s}^{-1}$ giving an interaction rate with a spin-polarized sphere of size $R$:

\begin{equation}
 \Gamma_\text{solar}\sim \frac{1}{\text{2.5 hours}}\left(\frac{n_s}{3\cdot 10^{22}~\text{cm}^{-3}}\right)^2\left(\frac{R}{10~\text{cm}}\right)^4.
\end{equation}

 Man-made neutrinos similar to those of the sun are a benchmark of the development of nuclear technologies. Taking as a benchmark a commercial nuclear power plant isotropic flux of roughly $10^{21}\frac{\bar{\nu}}{\text{sec}}$, we find an interaction rate of:
\begin{equation}
     \Gamma_\text{reac}\sim \frac{1}{\text{3 hrs}}\left(\frac{n_s}{3\cdot10^{22}~\text{cm}^{-3}}\right)^2\left(\frac{R}{10~\text{cm}}\right)^4\left(\frac{100~\text{m}}{d}\right)^2,
\end{equation}
where $d$ is the distance of the detector from the reactor neutrino source. This cross-section is independent of the relativistic neutrino energy. The maximum possible excitation energy in this case is $2 \times 10^{-6}\frac{10~\text{cm}}{R}$~eV, in order for the neutrinos to couple coherently to the entire structure.

Given these rates on such small detector, one can envisage the possibility of monitoring not only reactor activity, but nuclear tests around the globe as well.
 
Any nuclear weapon's test around the world will produce a huge burst of neutrinos or antineutrinos depending on the type of process that produces them. With little knowledge of the details of these processes, we make an estimate for the number of events expected from such a burst based on reactor parameters and we get:

\begin{equation}
 N_\text{bomb}\sim \mathcal{O}(1)\left(\frac{n_s}{3\cdot 10^{22}~\text{cm}^{-3}}\right)^2\left(\frac{R}{10~\text{cm}}\right)^4\left(\frac{10~\text{km}}{d}\right)^2 ,
\end{equation}
 for a $1$~Mton bomb.

 We stress that, while the rates may appear encouraging, they will only become useful when a concrete detection protocol is in place~\cite{Onurprotocol}. 
It is also worth mentioning that the characteristic energy of neutrinos is orders of magnitude above the characteristic $\omega_0$ of the systems we are considering. A successful protocol points to the possibility of ultra-low threshold detectors monitoring extremely high energy processes.

 \subsection{Axions sourced from Black Hole superradiance}
 \label{sec:SR-axions}

 Now we turn to astrophysical sources of BSM particles. One particularly loud example is axions sourced by black hole (BH) superradiance (SR).
 BHSR is a instability that affects rotating astrophysical black holes  when a boson has Compton wavelength close to the BH size. When the SR condition is satisfied, a cloud of bosons grows in a bound orbit around the BH, extracting a significant fraction of the BH's mass and angular momentum, forming a gravitational atom in the sky~\cite{axiverse,Arvanitaki-Dubovsky-BHSR-pheno}. BHSR is a very different from Dicke SR, they only share the same name, and they should not be confused.
 
 Axions with masses $10^{-13}$~eV$-$ $10^{-11}$~eV that exhibit strong self-interactions, $f_a\lesssim 10^{16}$~GeV, are emitted through energy exchange processes in the superradiant cloud, with energy density on Earth~\cite{BHSR-self-interactions} given by:

\begin{equation}
    \rho_\text{SR}\approx 10^{-6} \frac{\text{GeV}}{\text{cm}^3}\pare{\frac{\alpha}{0.1}}^6\pare{\frac{10\text{kpc}}{r}}^2\pare{\frac{f_a}{10^{16}\text{GeV}}}^2,
\end{equation}
where $\alpha\equiv G_NmM_\text{BH}$, $G_N$ is Newton's constant, $M_\text{BH}$ the mass of the BH, and $r$ the distance to the BH. Their velocity is set by the cloud dynamics and is $\alpha/6$ for fast spinning BHs that build up the lowest-lying superradiant levels. Because their masses are very small and the radiation very narrow-band, their coherence time is much longer compared to that of axion DM, and we can estimate the probability to induce precession in $\ket{D}$ using Eq.~\eqref{eq:Rabi_axion_Dicke}. For a BH at the opposite side of the Galaxy these are

\begin{equation}
    \begin{split}
        p_+=p_-&\approx 40\% \pare{\frac{N}{6\cdot 10^{17}}}^2\pare{\frac{\alpha}{0.1}}^8\pare{\frac{30\text{kpc}}{r}}^2\pare{\frac{t}{1\text{s}}}^2,
        \label{eq:Rabi_axion_SR}
    \end{split}
\end{equation}
where we have ignored precise details of the signal and its directionality, as well as the possibility of having a stochastic background of these axions from the entire Galactic BH population. This demonstrates that, ultimately, very small detectors could have the potential to be sensitive probes of quite diverse beyond the Standard Model physics and will not be restricted to cosmological relics. 

\section{Dark Quantum Optics for Cosmic Noise}
\label{sec:dark-quantum-optics}

While all the computations in the previous sections predict extremely large interaction rates even in small targets, it is not immediately obvious how this information can be read out. Traditional particle detector methods rely on energy exchange, which by definition is only sensitive to net interaction rates. These are generally suppressed and even zero for the case of stimulated absorption or emission.

In this section we show how we can go beyond energy exchange by thinking of the cosmic backgrounds as a bath of particles introducing \emph{noise} to our quantum detector. We emphasize, however, that this is \emph{tunable} noise with sundry control parameters: system size, geometry, orientation, and tuning energy, to name a few. In other words, the shape of the noise spectrum is at the hands of the experimentalist. 
This interpretation suggests treating the quantum detector as an open system and considering observables that encompass diffusion and decoherence, both of which have been extensively studied in the field of Quantum Optics.

This section is structured as follows: we first introduce the Lindblad formalism for open quantum systems and discuss what it reduces to for the C$\nu$B, DM, and other particle backgrounds. Then, we go on to contrast candidate energy exchange observables with diffusion and decoherence in these systems.

\subsection{The Lindblad equation}
\label{sec:lindblad}

The interaction of a quantum system with degrees of freedom that cannot be fully accounted for is nicely described by the  Gorini-Kossakowski-Sudarshan-Lindblad master equation:

\begin{eqnarray}
\dot{\rho}_S=-i[H_I,\rho_S]+\sum_A \gamma_A\mathcal{L}_{A}[\rho_S(t)],
\label{eq:lindblad}
\end{eqnarray}
where the Lindblad superoperator is $\mathcal{L}_{A}[\rho]\equiv2A\rho_S A^\dagger-A^\dagger A \rho_S-\rho_S A^\dagger A$, and $\gamma_A$ is the interaction rate of the bath that has been traced out with the system. The Lindblad equation assumes that the bath is Markovian in nature. For completeness, we point out that $H_I$ includes interactions with any degrees of freedom that still allow for unitary evolution. This formalism is crucial for quantum optics as it captures any energy exchange or loss of coherence of the quantum system. Spontaneous emission for example is easily included in those terms. 

Both in Eq.~\eqref{eq:lindblad} above and in what follows, all equations involving the density matrix are in the interaction picture for the system, whose free system Hamiltonian is $H_S=\omega_0 J_z$.

\subsubsection{The Markovian regime for the \texorpdfstring{C$\nu$B}~, DM and late-universe particles}
\label{sec:master_markov}

In our case, the properties of the C$\nu$B, Dark Matter in the Markovian regime as well as man-made neutrino sources can be captured and quantified by Eq.~\eqref{eq:lindblad}. The relevant operators for the inelastic processes we have already discussed are the raising and lowering collective operators $J_-$ and $J_+$. This treatment is exact only in the forward direction of the Rayleigh-Gans regime of scattering that dominates the scattering rate, and as such the Lindblad formalism remains a good approximation when describing the evolution of extended quantum systems. This point is further explained in app.~\ref{app:general}. When interactions seize to be coherent on the whole object, we can no longer describe the system in terms of collective Lindblad operators and the full Eq.~\eqref{eq:full_q} must be used.

The interaction rates $\gamma_+$ and $\gamma_-$ are the individual two-level system scattering rates as they have been defined in sections~\ref{sec:cnb}, and~\ref{sec:DM}. It is worth pointing out that for the C$\nu$B and DM scattering, the expectation value of the interaction Hamiltonian in the cosmic bath can be non-zero, $\avg{H_I}\neq 0$. It can be shown that in the RWA (Rotating Wave Approximation) only a term $\propto J_z$ survives. In an isotropic bath, this term averages out. If isotropy is broken due to a relative velocity between the bath rest frame and the system, this term gives rise to an energy shift  $\delta \omega_S$ proportional to $G_f v_{\text{rel}}(n_\nu-n_{\bar{\nu}})$ for the C$\nu$B, and $\Lambda^{-2} v_{\text{rel}}(n_{DM}-n_{\overline{DM}})$ for DM. The difference in densities denotes the particle-antiparticle asymmetry. For the C$\nu$B this is set in the SM by the baryon asymmetry, and this term is  what is known in the literature as the Stodolsky effect~\cite{stodolsky1975speculations}. There is also a similar term for man-made neutrino sources. The Lindblad equation becomes for the C$\nu$B, DM, and man-made neutrino sources:
\begin{equation}
    \begin{split}
        \dot{\rho}_S\simeq -i \delta \omega_S [J_z,\rho_S]+\frac{\gamma_-}{2}\mathcal{L}_{J_-}[\rho_S(t)]+\frac{\gamma_+}{2}\mathcal{L}_{J_+}[\rho_S(t)],
        \label{eq:nu_lind}
    \end{split}
\end{equation}

The $N^2$ coherent enhancement effects as we will see in section~\ref{sec:exp-observables} arise from both $\mathcal{L}_{J_+}$ and $\mathcal{L}_{J_-}$ terms. Before we examine their properties in more detail, we have to examine how the non-Markovian axion DM regime differs from the general discussion above.

\subsubsection{The non-Markovian regime for axions and dark photons}

The master equations for axion DM in the non-Markovian regime can be derived from the Born approximation of the general density matrix evolution equation. We find:

\begin{equation}
\begin{split}
    \dot{\rho}_S=&-\int_0^t\di \tau \,g^{(1)}(\tau)e^{-i\omega_0\tau}\\
    &\times\big[J_-J_+\rho_S(t-\tau)-J_-\rho_S(t-\tau)J_+\\
    &\quad\quad-J_+\rho_S(t-\tau)J_-+\rho_S(t-\tau) J_+J_-\big] + \text{h.c.},
     \label{eq:ax_lind_full}
\end{split}
\end{equation}
where the function $g^{(1)}(\tau)$ is the appropriate two-point correlation function of the bosonic particle. For the axion, it is the two-point correlation function of field gradients, given by eq.~\ref{eq:axion-corr}. Because $g^{(1)}$ has both rotating and counter-rotating terms, in the rotating wave approximation (RWA) only one of the two contributes to each term in Eq.~\eqref{eq:ax_lind_full}.
At long times, the equation above reduces to the Lindblad equation described in~\ref{sec:master_markov}, with $\gamma_+=\gamma_-$, as computed in sec.~\ref{sec:DMabs}. 

The limit of very long coherence times allows for some simplification. Taking $t\ll \tau_c$ we get $g_a^{(1)}(\tau)\approx \rho v_0^2\cos(m\tau)/f_a^2$, where we identify the Rabi frequency $\Omega^2\equiv 2\rho_\text{DM}v_0^2/f_a^2$. Tuning the splitting to $\omega_0=m$, instead of the Markovian Lindblad master equation we find

\begin{equation}
\begin{split}
    \dot{\rho}_S=-\frac{\Omega^2}{4}\int_0^t\di \tau \,\big[&J_-J_+\rho_S(\tau)-J_-\rho_S(\tau)J_+\\
    &-J_+\rho_S(\tau)J_-+\rho_S(\tau) J_+J_-\big] + \text{h.c.},
     \label{eq:ax_Rabi_1}
\end{split}
\end{equation}
which can be further simplified to

\begin{equation}
    \ddot{\rho}_S=\frac{\Omega^2}{4}\parea{\mathcal{L}_{J_-}[\rho_S(t)]+\mathcal{L}_{J_+}[\rho_S(t)]},
    \label{eq:ax_Rabi}
\end{equation}
with the imposed initial condition $\dot{\rho}_S(0)=0$. There is a connection between Eq.~\eqref{eq:ax_Rabi} and the Bloch equations. Computing the time-evolution of $J_z$, for instance, yields $\ddot{\avg{J_z}}=-\Omega^2\avg{J_z}$. While this reminds of Rabi oscillations, it is qualitatively different, because the initial condition for $\dot{\rho}$ sets $\avg{\dot{J_z}(0)}=0$. In constrast, in standard spin precession under a coherent field, this initial condition depends on the expectation of the spin in the $x$ and $y$ directions, through the Bloch equations, and can be non-zero. The difference arises because Eq.~\eqref{eq:ax_Rabi} is averaged over the phase or the direction of the bosonic field, both of which are zero-mean random variables in an isotropic background. We elaborate further on the link between non-Markovian axion noise and Rabi oscillations in App.~\ref{app:product_classical_axion}. A key point here is that axions with long coherence times can still be described as noise that induces non-Markovian evolution of the density matrix. 

Based on the formalism described above we are now ready to describe possible experimental observables.


\subsection{Towards measuring \texorpdfstring{$N^2$}~~effects}
\label{sec:exp-observables}

In this subsection, we discuss possible experimental observables that will be sensitive to the total interaction rate $\Gamma_{\text{total}}$ not just the net interaction rate $\Gamma_{\text{net}}$ and comment how the $N^2$ effects computed in Secs~\ref{sec:cnb},~\ref{sec:DM} and~\ref{sec:misc} can be read out of realistic systems, where we also take into account the variance of the state that is being studied. We will first discuss observables sensitive to energy exchange and then contrast them with observables that arise when we consider the coherent particle interactions as noise. This will naturally point us to observables that involve diffusion and decoherence.

We will calculate the signal-to-noise ratio (SNR) for these toy observables assuming that it is shot noise limited, and discuss which ones show a metrological advantage, i.e. the SNR scales favorably with powers of $N$. We expect that the main background stemming from spontaneous emission of photons can be suppressed through cavity engineering. For example, in a $V=1~\text{m}^3$ cavity of $Q=10^8$, the spontaneous emission rate of nuclear spins that are operated much below the cavity resonance is given by $\Gamma_S=\frac{N^2 g_N^2}{16 \pi^4 Q V}= 2\times 10^{-4}~\text{Hz}\pare{\frac{N}{10^{15}}}^2$~\cite{Lukinlectures}. We take $g_N$ to be the gyromagnetic ratio of the neutron. For the parameters chosen, cavity dissipation is sufficient, but this no longer is the case when $N$ becomes very large, as probably dictated for the case of the C$\nu$B. Addressing this regime goes beyond the scope of this paper, as what we present here is intended as ``proof of principle". 

We neglect additional sources of noise that come from the spin relaxation times $T_1$ and $T_2$~\cite{slichter2013principles}. When cooperative decay is suppressed, these timescales are set by \emph{local} effects and they do not appreciably change the states as long as the experiment is run for times shorter than these timescales. We will present the details of these and all additional sources of noise in upcoming work, where a concrete protocol will be proposed~\cite{Onurprotocol}.

\subsubsection{Energy exchange}

Traditional particle detection experiments, such as DM direct detection or axion detection setups rely on energy exchange. For a two-level system, a direct measure of energy transfer is $\langle J_z\rangle\equiv \text{tr}(\rho_S J_z)$, since the Hamiltonian is simply $H=\omega_0 J_z$. Working in perturbation theory, we will assume $\rho_S \approx \rho_S(t=0)$ in the Lindblad superoperators. Equation~\ref{eq:nu_lind} becomes, ignoring any effects from $\delta \omega_S$:

\begin{eqnarray}
    \rho_S(t)\approx \frac{\gamma_- t}{2}\mathcal{L}_{J_-}[\rho_S(0)]+\frac{\gamma_+ t}{2}\mathcal{L}_{J_+}[\rho_S(0)],
\end{eqnarray}
in the Markovian regime, and:
\begin{eqnarray}
    \rho_S(t)\approx \frac{\Omega^2 t^2}{2}\mathcal{L}_{J_-}[\rho_S(0)]+\frac{\Omega^2 t^2}{2}\mathcal{L}_{J_+}[\rho_S(0)],
\end{eqnarray}
in the deeply non-Markovian regime for ultralight DM. Perturbation theory will break down when there is $\mathcal{O}(1)$ fractional change in $\rho_S(t)$, or equivalently for an arbitrary spin observable $\mathcal{O}$, perturbation theory applies as long as the correction $\avg{\mathcal{O}(t)}-\avg{\mathcal{O}(0)}\ll\max[\avg{\mathcal{O}(0)},1]$.

With this in mind, we are in a position to calculate the value of $\langle J_z\rangle$ for three initial states of interest: $\ket{G}$, $\ket{P}$, and $\ket{D}$. The results of our calculation are summarized in Table~\ref{tab:jz}. 

For $\ket{D}$, intuition about the scalings of Table~\ref{tab:jz} can be gained by solving for the density matrix directly:

\begin{table*}[ht!]
    \centering
    \begin{tabular}{|c||c|c|c||c|}
    \multicolumn{5}{c}{Energy Exchange} \\
    \hline\multicolumn{4}{|c||}{C$\nu$B, scattering \& general Markovian noise} & Non-Markovian Noise\\\hline\hline
    $\avg{\mathcal{O}(t)}$ & $\ket{G}$ & $\ket{P}$ & $\ket{D}$ & $\ket{G},\ket{P},\ket{D}$\\\hline
        $\avg{J_z(t)}$ & $-\frac{N}{2} + N \gamma_+ t$ & $\frac{N^2}{4}(\gamma_+-\gamma_-)t$ & $\frac{N^2}{4}(\gamma_+-\gamma_-)t$ & \\
        $\Delta J_z$ & $N\gamma_+ t$ & $N/4$ & $\frac{N^2}{4}(\gamma_++\gamma_-)t$ & $\gamma_\pm t\to\Omega^2t^2/2$\\
        SNR & $\sqrt{N\gamma_+ t}$ & $\frac{N^{3/2}}{2}(\gamma_+-\gamma_-)t$ & $\parea{\frac{N^2}{4}\frac{(\gamma_+-\gamma_-)^2}{\gamma_++\gamma_-}t}^{1/2}$ & \\\hline
    \end{tabular}
    \caption{Value of average spin-flips, $\avg{J_z}$, or energy exchange, $\avg{H}=\omega_0\avg{J_z}$, as a function of time for three different initial states for a system of spins interacting with the C$\nu$B, DM or late-universe neutrinos and axions. We give only the relevant leading powers of $N$. The SNR here is defined as $\parea{\avg{J_z(t)}-\avg{J_z(0)}}/\sqrt{\Delta J_z}$ and is meant to represent the fundamental intrinsic noise of the system. \emph{Left:} Markovian cosmic backgrounds. The notation $\gamma_\pm$ refers to the coefficients of the Lindblad superoperators $\mathcal{L}_{J_\pm}$, respectively. They are related with $\Gamma_{\text{total}(\text{net})}=N^2 (\gamma_++(-)\gamma_-)$. In particular, the net rate is exactly zero for absorption/emission-type interactions. \emph{Right:} If the noise is in the deep non-Markovian regime, with coherence time $\tau_c\gg t_\text{exp}$, and to leading order in time, the $N$ scalings are the same. The corresponding time-evolution can be read off immediately from those on the left part of the Table, with the substitution $\gamma_\pm t \to\Omega^2 t^2/2$, where $\Omega$ is the Rabi frequency. Subtleties of the non-Markovian computation are addressed in App.~\ref{app:product_classical_axion}.}
    \label{tab:jz}
\end{table*}

\begin{equation}
\begin{split}
    \rho(t)\approx&\parea{1-p_-(t)-p_+(t)}\ket{\frac{N}{2},0}\bra{\frac{N}{2},0}\\
    &+p_-(t)\ket{\frac{N}{2},-1}\bra{\frac{N}{2},-1}+p_+(t)\ket{\frac{N}{2},1}\bra{\frac{N}{2},1},
    \label{eq:rho_D}
\end{split}
\end{equation}
where

\begin{equation}
    \begin{split}
            p_+(t)&=p_-(t)=2\int_0^t\di t'\int_0^{t'}\di\tau\, g^{(1)}(\tau) \cos\omega_0 \tau,\\&\qquad\qquad\qquad\qquad\qquad\qquad\text{absorption/emission,}\\
    p_+(t)&=\frac{N^2}{4}\gamma_+ t,\quad p_-(t)=\frac{N^2}{4}\gamma_- t,\\&\qquad\qquad\qquad\qquad\qquad\qquad\text{scattering},
    \label{eq:rho_D_sol}
    \end{split}
\end{equation}
to leading order in $N^2$. The expression labeled as ``scattering'' assumes Markovian evolution under scattering-type interactions, which is the case for the neutrinos and the fermionic DM considered in this work. The expression labeled as ``absorption/emission'' applies for arbitrary coherence times for interactions exhibited by the SM photon, the axion and the dark photon, neglecting spontaneous emission. For the axion, for instance, one finds

\begin{equation}
\begin{split}
    p_+(t)=p_-(t)=&
    \frac{2\rho_\text{DM}N^2}{3\sqrt{\pi}v_0^3 m^5 f_a^2}\int\di k\, k^4 e^{-\frac{k^2}{m^2v_0^2}}\\
    &\times\parea{\frac{\sin\pare{m+\frac{k^2}{2m}-\omega_0} t/2}{\pare{m+\frac{k^2}{2m}-\omega_0}/2}}^2,\quad\text{axion},
    \label{eq:axion_probabilities_D}
    \end{split}
\end{equation}
where we have applied the RWA.

We note, incidentally, that perturbation theory breaks down for a Dicke state already at times $t\gtrsim (N^2\gamma_\pm)^{-1}$. 

This solution illustrates that the cancellation of the $N^2$ effects as suggested by Table~\ref{tab:jz}, partial or total, depends on the question we ask. If we inquire, for instance, ``What is the probability that a measurement of $J_z$ will yield $+1$ after time $t$?'', we get the immediate answer: ``$p_+(t)$'', which scales as $N^2$. This also means that starting from the Dicke state $\ket{D}$ allows for measuring the excitation and de-excitation rates independently. Then  for $\ket{D}$, a signal-to-noise ratio (SNR) of $\mathcal{O}(1)$ can be achieved in the time it takes to get one event, i.e. $t\sim (N^2 \gamma_\pm)^{-1}$, a fact that is not captured by the naive observable $\avg{J_z}$.

Furthermore, we see that starting from $\ket{G}$ and $\ket{P}$, the SNR does not grow as fast, as summarized in Table~\ref{tab:jz}. For $\ket{G}$ the scalings are well-known, but for $\ket{P}$ we see the complete cancellation of $N^2$ effects when the net rate is zero. Here it is harder to find an analogue of the probability to excite or de-excite the system, as we did in the previous paragraph for $\ket{D}$, because $\ket{P}$ is intrinsically noisy. Interactions still occur, however, so, again, we see that a different observable is needed.

Nevertheless, energy exchange observables illustrate two important points. First, from discussion regarding $\ket{D}$, a significant metrological advantage can be gained even when there is no net energy exchange. 
Second, the SNR for the $\ket{P}$ state grows with $N^{3/2}$. Comparing when the SNR is $\mathcal{O}(1)$ for the $\ket{P}$ and $\ket{G}$ states, we find that the $\ket{P}$ state offers a metrological advantage compared to the $\ket{G}$ state, when $\sqrt{N}\gamma_\text{net}/\gamma_+>1$. For the macroscopic samples used in axion DM detection experiments such as CASPEr~\cite{Casper-theory,Casper-proposal} or QUAX~\cite{Barbieri:1985cp,BARBIERI2017-QUAX}, if they are prepared and observed in the $\ket{P}$ state, they have a sensitivity to a local C$\nu$B density which is a factor of:
\begin{equation}
\begin{split}
C_\text{boost}\sim 2\cdot 10^{11}&\pare{\frac{10~\text{cm}}{R}}^{3/2} \pare{\frac{3\cdot 10^{22}~\text{cm}^3}{n_s}}^{3/2}\\
&\times\pare{\frac{1000~\text{sec}}{t}}\pare{\frac{10^3}{N_\text{shots}}}^{1/2},
\end{split}
\end{equation}
larger than the SM prediction. In the equation above, we
use the C$\nu$B rates as calculated in sec.~\ref{sec:nu-elastic}.
While such values of $C_\text{boost}$ are completely unphysical, they are comparable to the ones achieved with current state-of-the-art neutrino experiments~\cite{KATRINpaper}. Observables beyond energy exchange offer obvious improvements and this is what we discuss next.

\subsubsection{Diffusion and Decoherence}
In this section we discuss observables that do not rely on net energy exchange. These have to do with variances of observables and decoherence, which naturally arise when we think of the cosmic backgrounds as noise. We emphasize once more, however, that the sensitivity to the cosmic noise depends on experimental parameters: energy splittings, geometry, system sizes and densities, to name a few. While we look for noise, this is a \emph{tunable} noise.

\emph{Diffusion.} The expectation value of $\avg{J_z}$ describes how many spin flips occur \emph{on average}. This can be thought as a \emph{drift}, which may be zero or not, as seen in the previous section. The response of the system, however, should also be sensitive to \emph{fluctuations} of the cosmic noise. Intuitively, the bath particles flip spins randomly and induce a random walk in spin space. Therefore, the \emph{variance} of observables is expected to scale with the total interaction rate. This can be thought as \emph{diffusion} in spin space, to borrow the terminology of random walks.

\begin{table*}[ht!]
    \centering
    \begin{tabular}{|c||c|c|c||c|}
    \multicolumn{5}{c}{Diffusion} \\
    \hline\multicolumn{4}{|c||}{C$\nu$B, scattering \& general Markovian noise, $\Gamma_\text{tot}\equiv\Gamma_++\Gamma_-$} & Non-Markovian Noise\\\hline\hline
    $\avg{\mathcal{O}(t)}$ & $\ket{G}$ & $\ket{P}$ & $\ket{D}$ & $\ket{G},\ket{P},\ket{D}$\\\hline
        $\avg{J_z^2(t)}$ & $\frac{N^2}{4}-(N^2-N) \gamma_+ t$& $\frac{N}{4} + \frac{N^2}{4}\gamma_\text{tot}t$ & $\frac{N^2}{4}\gamma_\text{tot}t$ & \\
        $\Delta J_z^2$ & $N^3\gamma_+ t$ & $N^2/8$ & $\frac{N^2}{4}\gamma_\text{tot}t$ & $\gamma_\pm t\to\Omega^2t^2/2$\\
        SNR & $\sqrt{N\gamma_+ t}$ & $\sqrt{2}N\gamma_\text{tot}t$ & $\parea{\frac{N^2}{4}\gamma_\text{tot}t}^{1/2}$ & \\\hline
    \end{tabular}
    \caption{Time evolution of the RMS spin-projection in the $z$-direction, $\avg{J_z^2}$, in perturbation theory (see text). We give only the relevant leading powers of $N$. The SNR here is defined as $\parea{\avg{J_z^2(t)}-\avg{J_z^2(0)}}/\sqrt{\Delta J_z^2}$ and is meant to represent the fundamental intrinsic noise of the system. \emph{Left:} Markovian cosmic backgrounds. The notation $\gamma_\pm$ refers to the coefficients of the Lindblad superoperators $\mathcal{L}_{J_\pm}$, respectively. They are related with $\Gamma_{\text{total}(\text{net})}=N^2 (\gamma_++(-)\gamma_-)$. In particular, the net rate is exactly zero for absorption/emission-type interactions.  \emph{Right:} If the noise is in the deep non-Markovian regime, with coherence time $\tau_c\ll t_\text{exp}$, and to leading order in time, the $N$ scalings are the same. The corresponding time-evolution can be read off immediately from those on the left part of the Table, with the substitution $\Gamma_\pm t \to\Omega^2 t^2/2$, where $\Omega$ is the Rabi frequency. Subtleties of the non-Markovian computation are addressed in App.~\ref{app:product_classical_axion}.}
    \label{tab:jz2}
\end{table*}

Random-walk effects famously appear in quadratic observables. We thus compute $\avg{J_z^2}$, its variance and the corresponding SNR, and summarize the results in table~\ref{tab:jz2}. 

One immediately sees that, for both $\ket{D}$ and $\ket{G}$, $\Delta J_z^2=0$ at time $t=0$. This is not surprising as both of these states are energy eigenstates. Despite that, $\ket{D}$ proves superior as the SNR grows with $N$ instead of $\sqrt{N}$, which is the case of $\ket{G}$. The state $\ket{P}$ is now sensitive to $\gamma_{\text{total}}$, but offers no metrological advantage compared to the more traditional $\ket{G}$ state. 

Nevertheless, the margin to gain by squeezing $\ket{P}$ now becomes enormous, as $\avg{J_z^2}$ is sensitive to the sum of absorption and emission of ultralight particles, contrary to $\avg{J_z}$. Squeezing by a factor of $\xi$, i.e. for $\Delta J_z=N/\xi$, improves the SNR for $\avg{J_z^2}$ by the same factor. The natural next step then is to devise concrete experimental protocols that employ squeezing and target diffusion observables such as $\avg{J_z^2}$. This we address in upcoming work~\cite{Onurprotocol}. We note that the observables presented here and the quantum protocols that can be devised to measure them can be different from other squeezing experiments, such as the pioneering work in~\cite{Hosten:2016yho}, or from previous spin squeezing proposals in NMR experiments for axion DM~\cite{sushkovthesis}, more specifically.

\emph{Decoherence.} Another obvious consequence of tracing out the particle environment from the description of the quantum system is the loss of quantum coherence and entanglement information as time elapses. Intuitively this occurs because background particles interact with the system, get entangled with it and then fly away undetected. This phenomenon is known as decoherence, which renders the system more classical over time. 

While decoherence plagues quantum metrology, here it can serve as an observable, paving the way to ultralow threshold detectors relying on loss of quantum correlations, rather than energy transfer. For example, the idea of anomalous, i.e. excess, decoherence as a probe of DM has been considered before in the context of \emph{elastic} scattering, relying on the loss of fringe contrast in atom interferometers~\cite{Riedel:2012ur,Riedel:2016acj,Du:2022ceh}. Similarly, appropriate observables could be devised for (pseudo)spin systems. We do not commit to such a one here, but only illustrate the loss of quantum correlations.

Although not an observable per se, the \emph{purity}, defined as $\text{tr}[\rho_S^2(t)]$, provides a measure of the loss of quantum coherence~\cite{Petruccione}. If $\text{tr}[\rho_S^2(t)]=1$, the system is in a pure state, while if $\text{tr}[\rho_S^2(t)]<1$, the system is in a mixed state in which decoherence has occured. For Markovian noise and to first order in perturbation theory we find:

\begin{equation}
    \begin{split}
            \text{tr}[\rho_S^2(t)]\approx1&-t\gamma_-\parea{\avg{J_+J_-}-\avg{J_+}\avg{J_-}}\\
            &-t\gamma_+\parea{\avg{J_-J_+}-\avg{J_+}\avg{J_-}},
    \end{split}
\end{equation}
where $\gamma_\pm$ are the coefficients of the $\mathcal{L}_{J_\pm}$ operators for the appropriate particle and the $\avg{\dots}$ denotes averaging with respect to the initial state of the system. For ultralight particles in the deeply non-Markovian regime we find instead

\begin{equation}
    \text{tr}[\rho_S^2(t)]\approx1-\frac{\Omega^2t^2}{4}\parea{\avg{J_+J_-}+\avg{J_-J_+}-2\avg{J_+}\avg{J_-}},
\end{equation}
but we refer the reader to App.~\ref{app:product_classical_axion} for the subtleties in interpreting the non-Markovian computation.

\begin{table*}[ht!]
    \centering
    \begin{tabular}{|c||c|c|c||c|}
    \multicolumn{5}{c}{Decoherence} \\
    \hline\multicolumn{4}{|c||}{C$\nu$B, scattering \& general Markovian noise} & Non-Markovian Noise\\\hline\hline
    Purity Loss & $\ket{G}$ & $\ket{P}$ & $\ket{D}$ & $\ket{G},\ket{P},\ket{D}$\\\hline
        $1-\text{tr}[\rho^2_S(t)]$  & $N\gamma_+ t$ & $N(\gamma_++\gamma_-)t/4$ & $N^2(\gamma_++\gamma_-)t/2$ & $\gamma_\pm t\to\Omega^2t^2/2$\\\hline
    \end{tabular}
    \caption{Purity loss, $1-\text{tr}[\rho^2_S(t)]$, of different detector states to leading order in perturbation theory. While purity is not an observable per se, it points towards loss of quantum correlations as a possible observable.}
    \label{tab:rho2}
\end{table*}

We tabulate the results in  Table~\ref{tab:rho2}. Purity loss scales as $N^2$ for $\ket{D}$ and as $N$ for $\ket{G}$. For the product state, although both $\avg{J_\pm J_\mp}$ and $\avg{J_\pm}\avg{J_\mp}$ scale as $N^2$ individually, the difference scales only as $N$. Since $\ket{P}$ has no quantum mechanical correlations---indeed $\ket{P}$ is a \emph{product} state---adding into or removing from it a quantum of energy doesn't change the state too much, i.e. $J_+$ and $J_-$ are uncorrelated to leading order. Again, this points towards the need to squeeze $\ket{P}$ to a metrologically more useful quantum state. 

Squeezing thus ties diffusion and decoherence together as possible observables and underlines the enormous gain in sensitivity we may achieve by using detectors with even small amounts of spin-spin correlations, while simultaneously going beyond observables that measure energy exchange. These considerations point to a new class of ultra-low threshold detectors. Of course, as we have repeatedly stressed, the discussion above is not meant as an experimental proposal and explicit protocols need to be designed to extract the maximum possible sensitivity of coherent inelastic processes on extended quantum systems. A first step towards this will be presented in upcoming work~\cite{Onurprotocol}.

\subsection{Decoherence due to elastic effects}
\label{sec:el_dephase}

Inelastic effects involve the projection of background operators (spin, polarization, momentum) in directions perpendicular to the spin of the detector: for spins polarized along the $z$-direction these are described by the operators $J_\pm$. Nevertheless, there are terms in all Hamiltonians $\propto J_z$ which we have largely neglected thus far. Similarly to inelastic effects, the Lindblad equation in general also contains $\mathcal{L}_{J_z}$:

    \begin{equation}
    \begin{split}
        \dot{\rho}_S\simeq -i \delta \omega_S [J_z,\rho_S]+\frac{\gamma_-}{2}\mathcal{L}_{J_-}[\rho_S(t)]&+\frac{\gamma_+}{2}\mathcal{L}_{J_+}[\rho_S(t)]\\
        &+\frac{\gamma_z}{2}\mathcal{L}_{J_z}[\rho_S(t)].
        \label{eq:nu_lind_deph}
    \end{split}
\end{equation}

The last term describes \emph{elastic} effects where no energy exchange with the system occurs; indeed $J_z$ trivially commutes with the free Hamiltonian of the spin system. Such terms are well-known in studies of \emph{decoherence}~\cite{Petruccione} as they describe a purely quantum-mechanical effect: the loss of quantum correlations without any energy dissipation. Since this work is focused on inelastic effects, we only discuss elastic effects for completeness and leave an exhaustive study to future work.

Because of the elastic nature of these effects, they do not appear at the same order as $\gamma_\pm$ for the absorption/emission-type interactions of axion, dark photons or ordinary photons; instead particle number conserving processes are the only relevant ones. Thus, for neutrinos and DM that scatters, the rate $\gamma_z$ is of the same order as $\gamma_\pm$. For instance, for elastic C$\nu$B neutrino scattering we find

\begin{equation}
    \gamma_z=n_\nu\frac{2\gf^2|u_{t,i\to i}|^2m_\nu^2 }{\pi}\avg{v_\nu},
    \label{eq:gamma_z}
\end{equation}
where the coefficients $u_{t,i\to i}$ are given by the appropriate entries of Table~\ref{tab:u-coeff}. As with all rates in the Lindblad equations, Eq.~\eqref{eq:gamma_z} is strictly correct in the limit $|\mathbf{q}|^{-1}\ll R$, but the Rayleigh-Gans regime can still be qualitatively captured by an analogous approximation as that described in App.~\ref{app:general}. 

For $\ket{D}$ we can immediately see that this new term does not change the solution Eq.~\eqref{eq:rho_D}. It is similarly easy to check that the contribution of this new term to $\avg{J_z^n}$ is zero for any $n$ and any state. Similarly, $\frac{\di}{\di t}\avg{J_x^n}=0$ and $\frac{\di}{\di t}\avg{J_y^n}=0$ for any $n$ in $\ket{D}$, but $\avg{J_x^2}=\frac{N^2}{4}-\frac{N^2}{4}\gamma_z t$ for $\ket{P}$, in the lab frame after averaging over many $\omega_0^{-1}$ periods. For this observable, it is the \emph{variance} that is of relevance for the SNR and we find $\Delta J_x=\frac{N}{4}-\frac{N^2}{4}\gamma_z t$, and again the time needed to have a signal above noise is $t\sim (\gamma_z N)^{-1}$, as in Table~\ref{tab:jz2}. 

From these scalings, it is clear that if one squeezes in energy, so that $\ket{P}$ approaches $\ket{D}$, the effect of $\gamma_z$ disappears from all spin observables. This is a manifestation of the fact that $\mathcal{L}_{J_z}$ induces phase uncertainty into the system. Because $\ket{D}$ already has infinite uncertainty in the phase, as an energy eigenstate, this term does not affect it. This justifies dropping this term thus far.

Instead, since $\ket{P}$ clearly has sensitivity to $\gamma_z$, we are naturally pointed towards a state with a well defined phase and increased uncertainty in the energy, i.e. the conjugate-variable analogue of $\ket{D}$, in order to read out these $N^2$ \emph{elastic} dephasing effects. This consideration is also hinted to by the fact that it is the variance in the $x$ and $y$ spin projections of $\ket{P}$ that are sensitive to $\gamma_z$, and not functions of $J_z$.

Let us take the NOON state as a concrete example, 
\begin{equation}
    \begin{split}
        \ket{\text{NOON}}&\equiv(\ket{N/2,-N/2}+\ket{N/2,+N/2})/\sqrt{2}\\
        &\equiv(\ket{G}+\ket{E})/\sqrt{2},
    \end{split}
\end{equation}
where we defined the fully excited state $\ket{E}\equiv \ket{N/2,+N/2}$ for brevity. The NOON state is extremely challenging to produce experimentally, and is used here only for illustration purposes because it is analytically tractable. While $\ket{\text{NOON}}$ has large phase uncertainty, it is not the same state as what one should get by squeezing $\ket{P}$ in phase, rather than energy. Since simple analytical arguments become more involved in the latter case, we defer it to future work.

Starting with $\rho=\ket{\text{NOON}}\bra{\text{NOON}}$ and considering the effects of $\mathcal{L}_{J_z}$ only, we have to first order in perturbation theory in time:

\begin{equation}
    \begin{split}
        \rho(t)\approx\ket{G}\bra{G}+\ket{E}\bra{E}&+\pare{1-\frac{N^2\gamma_z t}{2}}\ket{G}\bra{E}\\
        &+\pare{1-\frac{N^2\gamma_z t}{2}}\ket{E}\bra{G},
    \end{split}
    \label{eq:noon_sol}
\end{equation}
which should be contrasted to the solution for $\ket{D}$ in Eq.~\eqref{eq:rho_D} for the inelastic case. We see that correlations decay as $N^2$. Thus, observables that are sensitive to projections on $\ket{G}\bra{E}$ or $\ket{E}\bra{G}$ will pick up these $N^2$ effects.  Similarly to the case of $\ket{P}$ and $\ket{D}$ for inelastic processes, any state that is squeezed in phase should recover some enhanced metrology compared to $N$ by the squeezing factor, ultimately approximating a sensitivity similar to $\ket{\text{NOON}}$ for large squeezing.

As we have emphasized before, elastic dephasing effects occur for scattering-type interactions, which already seem to provide superior observables through their \emph{inelastic} effects. This is because, first, elastic dephasing still relies on some quantum manipulation of the system, such as squeezing, to be metrologically useful and, second, because inelastic effects have more experimental control parameters.

Nevertheless, we wish to emphasize that the elastic dephasing presented here may be of relevance. First, it demonstrates that pure decoherence effects are not restricted to loss of fringe contrast in atom-interferometers, which rely on localization of the center-of-mass~\cite{Riedel:2012ur,Riedel:2016acj,Du:2022ceh}. In such proposals the atoms are in superposition of positions and decoherence occurs because short-wavelength DM affects the two paths of the interferometer in an uncorrelated way. Here, decoherence occurs in the phase of a spin system and, similarly, it should appear in any setup where the target particles can be put in superpositions of states. This points to different experimental setups, even for the same cosmic background.

Second, the quantum states required for sensitivity to elastic effects differ significantly from those needed for inelastic effects. This implies that distinct quantum protocols would be necessary for each scenario. Additionally, this differentiation suggests that experimental noise will influence elastic and inelastic searches in varying ways. For instance, Dicke superradiance of photons must be mitigated for inelastic effects at very large $N$, whereas it is much less significant for dephasing effects: for instance $\mathcal{L}_{J_-}$ has only an $\mathcal{O}(N)$ impact on $\ket{\text{NOON}}$. Consequently, it is plausible that tailored quantum protocols could be developed to detect different effects of cosmic backgrounds, perhaps also aiding in signal discrimination.

\section{Discussion and outlook}
\label{sec:outlook}

In this paper, we show the conditions under which interactions scale like the square of the number of targets, while exciting the target's internal degrees of freedom. These superradiant interaction rates for cosmic relics as well as late-universe sourced particles can be quite large even for modest sized systems. The two-level systems we consider are composed of electron or nuclear spins not only for concreteness, but also because electron and nuclear spin resonance techniques have been extensively studied. Our results can be generalized to any two-level system and any coupling of interest.

The processes described above offer two obvious advantages. Firstly, the size of the detector required for detection can be greatly reduced. Our benchmark rates are always calculated for objects no larger than $R=10$~cm.

Secondly, the energy splitting $\omega_0$ that is required for superradiant interactions to be efficient can very small compared to the kinetic energy of the scattering particles. Detectors based on  superradiant interactions, if realized experimentally, will be in essence a new type of ultra-low threshold detector.  

Section~\ref{sec:dark-quantum-optics} shows that superradiant interactions can be thought of as noise in a quantum optics system. While we do not describe a concrete experimental setup that will be sensitive to the rates we calculate, protocols that are sensitive to diffusion or decoherence observables in combination with squeezing have the best perhaps chance of providing a viable setup where these effects can be detected. 

It is worth pointing out though that for scattering, traditional energy transfer observables can still be sensitive to superradiant effects. Due to the differences in phase space for excitation and de-excitation, the net rate can still be proportional to $N^2$. For the most interesting theoretically case of the C$\nu$B, this means that current experiments looking for axion DM may be able to place a comparable constraint on the C$\nu$B than the current leading neutrino experiments~\cite{KATRINpaper}, after a modest observation time as calculated in~sec.~\ref{sec:exp-observables}. Superradiant interactions can provide for improvements in searches already, without the need of specialized quantum metrology protocols.

Ultimately, though, in order to extract the most out of superradiant interactions of cosmic relics or weakly interacting particles, one will need to address the issue of experimental detection of the diffusion or decoherence such interactions imply. There are also control parameters that can help us distinguish the signal from systematics, such as the dependence on the geometry and size of the target, as well as the excitation energy $\omega_0$. The importance of these discriminators will have to be evaluated when a specific protocol is designed; this is what we plan to move to next~\cite{Onurprotocol}. 

\section*{Acknowledgements}

We are grateful to Onur Hosten, Mark Kasevich, Jason Hogan, Monika Schleier-Smith, Markus Aspelmeyer, David Schuster, Alex Sushkov, and Sisi Zhou for illuminating discussions. Over the last few years, we have been fortunate to share enlightening discussions about neutrino physics with Giorgio Gratta. We are also thankful to Junwu Huang, and Cristina Mondino for useful discussions on particle DM models. We finally thank Junwu Huang for feedback on a prepublication version of the paper. Research at Perimeter Institute is supported in part by the Government of Canada through the Department of Innovation, Science and Economic Development and by the Province of Ontario through the Ministry of Colleges and Universities. AA is grateful for the support of the Stavros Niarchos Foundation and the Gordon and Betty Moore foundation. SD acknowledges support by NSF Grant PHY-2310429 and the Gordon and Betty Moore Foundation Grant GBMF7946.

\appendix
\section{General spin states}
\label{app:general_states}

In this Appendix we compute the collective enhancement factor for the different processes described in the main text, using general spin coherent states and general Dicke states. While maximal rates are achieved for $\ket{P}$ in Eq.~\eqref{eq:P_Fourier} and the equatorial Dicke state $\ket{N/2,0}$, we include the generalized states here for completeness. We follow~\cite{Mandel_Wolf_1995} for the definitions below.

A general coherent spin state is defined in analogy to bosonic coherent states. The displacement operator for a spin system is defined as $D\equiv \exp(\zeta J_+-\zeta^* J_-)$ and can be factorized as follows

\begin{equation}
\begin{split}
    e^{\zeta J_+-\zeta^* J_-}&=e^{z J_+}e^{J_z\log\pare{1+|z|^2}}e^{-z^*J_-}\\
    &\equiv D(z),
\end{split}
\end{equation}
where $z\equiv \tan\abs{\zeta} e^{i\text{arg }\zeta}$. 

The general atomic coherent state $\ket{l,z}$ with  $0\leq l\leq N/2$, is defined as $\ket{l,z}=D(z)\ket{l,-l}$, where $\ket{l,-l}$ is the ground state of the atomic system in the subspace of a fixed $l$, i.e. in the subspace of a fixed \emph{total} angular momentum $J^2$. Then

\begin{equation}
   \ket{l,z}=\frac{1}{\pare{1+|z|^2}^{1/2}}\sum_{m=-l}^l 
   \begin{pmatrix}
       2l \\
       l+m
   \end{pmatrix}^{1/2}
   z^{l+m}\ket{l,m},
\end{equation}
where $\ket{l,m}$, with $m\in\left\{-l,...,l\right\}$ are the Dicke states of fixed total $l$.

The operator expectation values that are needed for all the processes we computed in this work are $\avg{J_\pm J_\mp}$, which on the state $\ket{l,z}$ are

\begin{equation}
\begin{split}
    \avg{J_+ J_-}_{\ket{l,z}}&=\frac{2l \abs{z}^2}{\pare{1+\abs{z}^2}^2}\pare{2l+\abs{z}^2},\\
    \avg{J_- J_+}_{\ket{l,z}}&=\frac{2l}{\pare{1+\abs{z}^2}^2}\pare{2l\abs{z}^2+1}.
\end{split}
\end{equation}

From this it becomes clear that all scattering rates are maximized for $l=N/2$, in which case all the coherent spin states $\ket{N/2,z}$ become product states, where each spin is in a pure superposition state and the spins are independent

\begin{equation}
    \ket{N/2,z}=\prod_{i=1}^N \frac{1}{\pare{1+\abs{z}^2}^{1/2}}\pare{\ket{g}+z\ket{e}}_i.
    \label{eq:general_P}
\end{equation}

These, in turn, give maximum rates for $\abs{z}=1$, which corresponds to the $\ket{P}$ state, Eq.~\eqref{eq:P_Fourier}, used in the main text. Equation~\eqref{eq:general_P} can be re-written in a simpler form as 

\begin{equation}
    \ket{\text{P}(c_g,c_e)}\equiv\prod_{i=1}^N \pare{c_g\ket{g}+c_e\ket{e}}_i,
    \label{eq:general_P}
\end{equation}
where $|c_g|^2+|c_e|^2=1$. Then it is simple to show that 

\begin{equation}
\begin{split}
    \avg{J_+ J_-}_{\ket{\text{P}(c_g,c_e)}}&=N |c_e|^2 + N(N-1)|c_g|^2|c_e|^2,\\
    \avg{J_- J_+}_{\ket{\text{P}(c_g,c_e)}}&=N |c_g|^2 + N(N-1)|c_g|^2|c_e|^2,
\end{split}
\end{equation}
which explicitly shows the requirement to have superpositions of the ground and excited states to achieve macroscopic coherence.

Similarly, for the scattering rates on Dicke states $\ket{l,m}$ we immediately find

\begin{equation}
\begin{split}
    \avg{J_\pm J_\mp}_{\ket{l,m}}=(l\pm m)(l\mp m+1),
\end{split}
\end{equation}
which are again maximum for $l=N/2$ and $m=0$.

These assume that the wavelength of the incoming or outgoing particles is longer than the system size. When this is no longer satisfied, one needs to use the generalized operators described in App.~\ref{app:general} below to derive the corresponding form factor.

\section{Lindblad equation for general momentum transfer}
\label{app:general}

All the Lindblad equations in the text assume that the background has a wavelength longer than the target, i.e. are correct in the limit $R\ll k^{-1}$, where $k$ the typical momentum scale of the incoming particles. For C$\nu$B neutrinos the size of the targets needs to be a few orders of magnitude larger than that to give appreciable interaction rates. This may apply to other cosmic backgrounds as well, depending on the total interaction rate. For fermionic DM, for instance, whose mass must necessarily be $\gtrsim$~keV and whose momentum is $\gtrsim$~eV, the condition $R\ll k^{-1}$ is never satisfied. Despite that, the coupling can still be coherent with the \emph{entire} target, with scattering happening predominantly in the forward direction. 

First we derive the generalized Lindblad equation for the C$\nu$B, taking into account the possible induced phases between the atoms that destroy the coherent coupling and comment on some aspects of it. Generalizing to any particle with an arbitrary coherence time is straightforward. We find\footnote{We emphasize that this equation applies to the particular helicity structure of the C$\nu$B. If the neutrino source is polarized then the appropriate polarized matrix elements should be computed. See Sec.~\ref{sec:misc} for some examples of such cases.}

\begin{widetext}
    \begin{eqnarray}
    \begin{aligned}
        \dot{\rho}_S(t)=&\frac{\gf^2|u|^2}{2}\int\frac{\di^3k}{(2\pi)^3}\int\frac{\di^3k'}{(2\pi)^3}\avg{n_k}(1-\avg{n_{k'}})\frac{4\pi(E+k)(E'-k'\cos\theta\cos\theta')}{EE'}\sum_{\alpha,\beta}e^{i\mathbf{q}\cdot\Delta\mathbf{r}_{\alpha\beta}}\\
        &\times\bigg[\pare{2J_+^{(\alpha)}\rho_S(t)J_-^{(\beta)}-J_-^{(\alpha)}J_+^{(\beta)}\rho_S(t)-\rho_S(t) J_-^{(\alpha)}J_+^{(\beta)}}\delta(E-E'-\omega_0)\\
        &+\pare{2J_-^{(\alpha)}\rho_S(t)J_+^{(\beta)}-J_+^{(\alpha)}J_-^{(\beta)}\rho_S(t)-\rho_S(t) J_+^{(\alpha)}J_-^{(\beta)}}\delta(E-E'+\omega_0)\bigg],
        \label{eq:full_q}
    \end{aligned}
\end{eqnarray}
\end{widetext}

where the sums are over the atomic positions $\mathbf{r}_\alpha,\mathbf{r}_\beta$, $\mathbf{q}\equiv\mathbf{k}-\mathbf{k}'$, $\Delta \mathbf{r}_{\alpha\beta}\equiv\mathbf{r}_\alpha-\mathbf{r}_\beta$ and the $J^{(i)}$ operators are the appropriate $2\times 2$ matrices for each individual spin of the target. Clearly, this equation reduces to Eq.~\eqref{eq:nu_lind} in the limit $(\mathbf{k}-\mathbf{k}')\cdot(\mathbf{r}_\alpha-\mathbf{r}_\beta)\ll1$ for every possible $\mathbf{k},\mathbf{k}',\alpha$ and $\beta$.

In the opposite limit of $\mathbf{q}\cdot\Delta\mathbf{r}_{\alpha\beta}\ll1$ the cross terms vanish and we are left with only one sum, so that instead of the two Lindblad superoperators of the total angular momentum we get $2N$ terms. Then the equation reduces to $\dot{\rho}_S\approx \frac{\gamma_-}{2}\sum _\alpha\mathcal{L}_{J_-^{(\alpha)}}[\rho_S]+\frac{\gamma_+}{2}\sum _\alpha\mathcal{L}_{J_+^{(\alpha)}}[\rho_S]$, which exhibits no coherent enhancement.

We can still write an effective description using collective spin operators, as long as the incoming neutrino couples coherently to the entire target. Again, we work out the the case of diagonal neutrino scattering. Generalizations to any other process or particle is straightforward.

We discretize the momenta $k$ and $k'$ and consider the expression

\begin{eqnarray}
\sum_\mathbf{k}\sum_{\mathbf{k}'}\sum_{\alpha,\beta}e^{i\mathbf{q}\cdot\Delta\mathbf{r}_{\alpha\beta}}J_+^{(\alpha)}J_-^{(\beta)}\delta(E-E'\pm\omega_0).
\end{eqnarray}

When $kR\gg 1$, the $\delta$-function imposes $|\mathbf{k'}|\approx|\mathbf{k}|$ for optimal tunings. Then $|\mathbf{q}|^2\simeq 2k^2(1-\cos\theta_{kk'})$, where $\theta_{kk'}$ is the relative angle between the two momenta. To couple coherently to the entire target we need $\mathbf{q}\cdot\Delta\mathbf{r}_{\alpha,\beta}\lesssim 1$, so $q_{\max}\approx 1/R$. This imposes the restriction $\theta_{kk'}\lesssim (kR)^{-1}$. Let us define as $\mathcal{C}_{\mathbf{k},\mathbf{k}'}$ the set of $\mathbf{k}'$ that satisfy this condition. Then we can break the sums into two terms

\begin{widetext}
    \begin{equation}
    \begin{split}
\sum_\mathbf{k}\sum_{\mathbf{k}'}\sum_{\alpha,\beta}e^{i\mathbf{q}\cdot\Delta\mathbf{r}_{\alpha\beta}}J_+^{(\alpha)}J_-^{(\beta)}&=\sum_\mathbf{k}\sum_{\mathbf{k}'\in\mathcal{C}_{\mathbf{k},\mathbf{k}'}}\sum_{\alpha,\beta}J_+^{(\alpha)}J_-^{(\beta)}+\sum_\mathbf{k}\sum_{\mathbf{k}'\notin\mathcal{C}_{\mathbf{k},\mathbf{k}'}}\sum_{\alpha,\beta}e^{i\mathbf{q}\cdot\Delta\mathbf{r}_{\alpha\beta}}J_+^{(\alpha)}J_-^{(\beta)}\\
&\approx\sum_\mathbf{k}\sum_{\mathbf{k}'\in\mathcal{C}_{\mathbf{k},\mathbf{k}'}} J_+J_- + \sum_\mathbf{k}\sum_{\mathbf{k}'\notin\mathcal{C}_{\mathbf{k},\mathbf{k}'}}\sum_{\alpha}J_+^{(\alpha)}J_-^{(\alpha)}
  \end{split}
\end{equation}
\end{widetext}
where we recover the collective operators in the first term, while the second term is, by construction, an incoherent sum for which the cross-terms vanish. 

The last term is incoherent and can be dropped to leading order. The condition $\mathbf{k}'\in \mathcal{C}$ only affects the kinematics and factorizes from the system operators. This allows for a Lindblad description exactly as in Eq.~\eqref{eq:nu_lind}, but with a modified rate. As long as there is enough energy to excite the atoms and couple to the entire target, the suppression comes only from the solid angle $\Delta \Omega'\approx(kR)^{-2}$. This recovers the Rayleigh-Gans scaling, as discussed in Sec.~\ref{sec:N2inel-intro}, at the level of the Lindblad equation.

\section{The non-Markovian regime and its connection to Rabi oscillations}
\label{app:product_classical_axion}

In this Appendix we explain some of the subtleties of the non-Markovian regime for ultralight particles, focusing on the axion case for concreteness. We discuss the behavior of $\ket{D}$ first, and then consider $\ket{P}$.

The state $\ket{D}$ is most instructive since we can solve for its time-evolution in perturbation theory. The solution, Eq.~\eqref{eq:rho_D_sol}, is repeated here simplified for convenience:

\begin{equation}
\begin{split}
    \rho(t)\approx&\parea{1-2p(t)}\ket{\frac{N}{2},0}\bra{\frac{N}{2},0}\\
    &+p(t)\ket{\frac{N}{2},-1}\bra{\frac{N}{2},-1}
    +p(t)\ket{\frac{N}{2},1}\bra{\frac{N}{2},1},
\end{split}
   \tag{\ref{eq:rho_D}}
\end{equation}
where

\begin{equation}
    \begin{split}
            p(t)\approx&\frac{2\rho_\text{DM}N^2}{3\sqrt{\pi}v_0^3 m^5 f_a^2}\\
            &\times\int\di k\, k^4 e^{-\frac{k^2}{m^2v_0^2}}\parea{\frac{\sin\pare{m+\frac{k^2}{2m}-\omega_0} t/2}{\pare{m+\frac{k^2}{2m}-\omega_0}/2}}^2.
    \end{split}
        \tag{\ref{eq:rho_D_sol}}
\end{equation}

For short coherence times the parametrics are $p(t)\propto N^2\Omega^2\tau_c t$, while for long coherence times $p(t)\propto N^2\Omega^2 t^2$.

The Lindblad treatment for axions predicts $\avg{J_z}=0$ for $\ket{D}$, as derived in Sec.~\ref{sec:exp-observables}. The solution~\eqref{eq:rho_D_sol} explains the cancelation: there is an equal probability that the system will be found in either $\ket{N/2,\pm1}$ after time $t$. To the extend that the axion is a stochastic field, these are classical probabilities arising after averaging over it.

Interpretating the averaging over the background field that is implicit in the Lindblad formalism, is straightforward in the Markovian regime, since absorption and emission events occur randomly. In the non-Markovian regime the system is influenced by a coherent field, whose phase and direction are a priori unknown. Because the Lindblad equation entails an average of the background over formally \emph{infinite} time, on average, after \emph{infinite} realizations of the experiment, the Lindblad solution simply says that the system will rotate towards $+z$ or $-z$ with the same probability.

One may worry whether this formalism is appropriate when the axion has a very long coherence time, much longer than the entire experimental program (years, centuries, Hubble$^{-1}$, etc). A particular example of this are the black hole superradiance axions, described in Sec.~\ref{sec:SR-axions}, whose coherence time can be as long as several years. In such cases, if the axion is detected, then its phase and direction can be measured. Then the direction of spin precession in the next experiment will be precisely known, an effect not captured by the Lindblad equation. In what follows we argue that, while the Lindblad formalism may be rendered inadequate \emph{after} the axion is detected, as far as only detecting the axion is concerned, this distinction is not parametrically important when moments of $J_z$ serve as observables, as described in Sec.~\ref{sec:dark-quantum-optics}.

To see this, we take the axion as a \emph{deterministic} driving perturbation, inducing Rabi oscillations. The Hamiltonian is

\begin{eqnarray}
    H=\frac{\Omega}{2}\pare{e^{-i\phi}J_++e^{i\phi}J_-},
    \label{eq:Rabi_H}
\end{eqnarray}
where $\Omega$ is a \emph{fixed} amplitude and $\phi$ a \emph{fixed} phase. Then the state after time $t$ becomes

\begin{equation}
\begin{split}
        \ket{\psi(t)}=&\pare{1-\frac{N^2\Omega^2 t^2}{16}}\ket{\frac{N}{2},0}\\
        &-i\frac{N\Omega t}{4}\pare{e^{-i\phi}\ket{\frac{N}{2},1}+e^{i\phi}\ket{\frac{N}{2},-1}}\\ &+\mathcal{O}(\Omega^2t^2),
\end{split}
    \label{eq:rho_Rabi_D}
\end{equation}
where we omitted a correction of order $\Omega^2t^2$ that involves the states $\ket{N/2,\pm2}$ for brevity.

While the density matrix $\rho'=\ket{\psi(t)}\bra{\psi(t)}$ is very different from $\rho$ of Eq.~\eqref{eq:rho_D}, the probability to be in any of the excited states $\ket{N/2,\pm 1}$ is still parametrically $p(t)\propto N^2\Omega^2 t^2$. The scaling of observables such as $\avg{J_z^2}$ also remains the same. Amusingly, $\avg{J_z}$ is zero in $\ket{D}$ even for a deterministic axion signal. We will see below that Rabi oscillations do induce a non-trivial time-dependence in $\avg{J_z}$ for $\ket{P}$.

The only difference between $\rho'$ and $\rho$ is that $\rho'$ describes a pure state, while the cross-terms, e.g. $\ket{N/2,0}\bra{N/2,\pm 1}$, have disappeared in $\rho$ due to decoherence; indeed, these terms are proportional to $\exp(\pm i
\phi)$, so after averaging over the uniformly distributed angle, they average to zero. In order to distinguish between the two density matrices, several measurements of the system must be carried out, possibly in different bases, with each measurement above noise. The signal-to-noise ratio in any of these measurements will be parametrically the same for both states, since the expectation values of observables are parametrically the same; the distinction will come about from $\mathcal{O}(1)$ effects. From a detection perspective then, whether the axion has an infinitely long coherence time (e.g. the superradiance axions), or one just as long as every shot of the experiment (e.g. QCD axion DM), is irrelevant. After detection, it is certainly true that the Lindblad and the Rabi predictions could be distinguished and provide us with a wealth of information about the axion and its origins. We leave the details of how this may be done to future work.

Now we turn to $\ket{P}$, to provide some further intuition on the non-Markovian regime. We also address in more detail the apparent loss of $N^2$ effects, as derived in Sec.~\ref{sec:dark-quantum-optics}.

Because $\ket{P}$ is simply the product of two-level systems, we can solve for its full time-evolution under the influence of the Hamiltonian~\eqref{eq:Rabi_H}. At time $t$ the state becomes

\begin{equation}
\begin{split}
    \ket{P(t)}=2^{-N/2}\prod_\alpha \bigg[&\pare{\cos\frac{\Omega t}{2}+ie^{i\phi}\sin\frac{\Omega t}{2}}\ket{g}\\
    &+\pare{\cos\frac{\Omega t}{2}+ie^{-i\phi}\sin\frac{\Omega t}{2}}\ket{e}\bigg].
\end{split}
\end{equation}

We may then compute the average spin in the $z$-direction, which yields

\begin{equation}
    \avg{J_z(t)}=\frac{N}{2}\sin\phi\sin\Omega t.
    \label{eq:jz_rabi_p}
\end{equation}

This result is markedly different from that of Table~\ref{tab:jz}, which was derived using the Lindblad equation and predicted $\avg{J_z}=0$. Averaging over the uniform $\phi$ in Eq.~\eqref{eq:jz_rabi_p}, which is what the Lindblad formalism implicitly does, trivially recovers the Lindblad result. 

One may then wish to compute the SNR using $\avg{J_z}$ as an observable. Because the variance $\Delta J_z$ of coherent spin states is $N$, we need time $t\approx(\sqrt{N}\Omega)^{-1}$ to be able to distinguish the state $\ket{P(t)}$ from the initial $\ket{P}$ and read out the interaction with the axion. One may understand the disappearance of the $N^2$ effects in this language: the state simply needs more time to evolve sufficiently above noise. The scaling of the required time is the same as the one suggested by Table~\ref{tab:jz2}, where the observable was $\avg{J_z^2}$. As in the case of $\ket{D}$, the state of the system is certainly different and distinguishing between the Lindblad and Rabi predictions can uncover valuable information about the axion signal. In terms of detection, however, the two approaches are equivalent.

\bibliography{refs}

\end{document}